\documentclass[acmlarge]{acmart}
\AtBeginDocument{%
  }

\makeatletter
\newcommand{\myconfshort}{\acmConference@shortname}
\newcommand{\myconffull}{\acmConference@name}
\newcommand{\myconfdate}{\acmConference@date}
\newcommand{\myconfloc}{\acmConference@venue}
\AtBeginDocument{
  \fancypagestyle{firstpagestyle}{
    \fancyhead{}%
    \fancyfoot[C]{}%
  }
  \fancyhf{}
  \fancyhead[LO]{\@headfootfont\shorttitle}%
  \fancyhead[RE]{\@headfootfont\@shortauthors}%
  \fancyhead[LE]{\@headfootfont\footnotesize \myconfshort, \myconfdate, \myconfloc}%
  \fancyhead[RO]{\@headfootfont\footnotesize \myconfshort, \myconfdate, \myconfloc}%
  \fancyfoot[C]{}%
}
\makeatother
\copyrightyear{2026}
\acmYear{2026}
\setcopyright{cc}
\setcctype{by}
\acmConference[FAccT '26]{The 2026 ACM Conference on Fairness, Accountability, and Transparency}{June 25--28, 2026}{Montreal, QC, Canada}
\acmBooktitle{The 2026 ACM Conference on Fairness, Accountability, and Transparency (FAccT '26), June 25--28, 2026, Montreal, QC, Canada}
\acmDOI{10.1145/3805689.3812329}
\acmISBN{979-8-4007-2596-8/2026/06}

\usepackage{enumitem}
\usepackage{listings}
\usepackage{cleveref}
\usepackage{soul}
\usepackage{color}
\usepackage{pifont}

\newcommand{\ie}{\textit{i.e.}}
\newcommand{\eg}{\textit{e.g.}}

\definecolor{lightgreen}{rgb}{0.85, 1.0, 0.85}
\sethlcolor{lightgreen}
\newcommand{\added}[1]{#1}
\newcommand{\baseline}{\textsc{Baseline}}
\newcommand{\outputUQ}{\textsc{UQ-Output}}
\newcommand{\relationUQ}{\textsc{UQ-Relation}}
\newcommand{\tokenUQ}{\textsc{UQ-Token}}

\newcommand{\radio}{\(\circ\)}
\newcommand{\checkbox}{\(\square\)}

\newcommand{\RadioScaleFive}{%
\begin{tabular}{@{}ccccc@{}}
\radio & \radio & \radio & \radio & \radio \\
{\footnotesize 1} & {\footnotesize 2} & {\footnotesize 3} & {\footnotesize 4} & {\footnotesize 5}
\end{tabular}
}

\newcommand{\LikertFive}{%
\begin{enumerate}[label=\radio, leftmargin=2.2em, itemsep=0.25em, topsep=0.25em]
  \item Strongly disagree (1)
  \item Disagree (2)
  \item Neither agree nor disagree (3)
  \item Agree (4)
  \item Strongly agree (5)
\end{enumerate}
}

\newcommand{\FreeResponseBox}[1][4.2cm]{%
\vspace{0.25em}
\par\noindent\fbox{\parbox{0.95\linewidth}{\vspace{#1}\hfill}}%
\par\vspace{0.25em}
}

\begin{document}

\title[Not All Uncertainty Is Equal]{Not All Uncertainty Is Equal: How Uncertainty Granularity Shapes Human Verification in LLM-Assisted Decision Making}


\author{Mauricio Villavicencio}
\email{villa669@umn.edu}
\orcid{0009-0007-6644-5810}
\affiliation{%
  \institution{University of Minnesota - Twin Cities}
  \city{Minneapolis}
  \state{Minnesota}
  \country{USA}
}

\author{Sitong Pan}
\email{pan00389@umn.edu}
\orcid{0009-0006-7729-6497}
\affiliation{%
  \institution{University of Minnesota - Twin Cities}
  \city{Minneapolis}
  \state{Minnesota}
  \country{USA}
}

\author{Qianwen Wang}
\email{qianwen@umn.edu}
\orcid{0000-0003-1728-4102}
\affiliation{%
  \institution{University of Minnesota - Twin Cities}
  \city{Minneapolis}
  \state{Minnesota}
  \country{USA}
}

\renewcommand{\shortauthors}{Villavicencio, Pan, Wang}

\begin{abstract}
  Despite ubiquitous warnings across major LLM platforms that these systems can make mistakes and need double check, users often develop inappropriate trust and accept incorrect answers without critical evaluation. Uncertainty quantification (UQ), \ie, displaying LLMs' confidence in their own responses, has emerged as a promising approach to calibrate user trust. 
  However, prior empirical studies on uncertainty communication have treated uncertainty as a single numerical score or simple natural language expression. 
  This simplification fails to capture a key property of LLM outputs: a single response often comprises multiple facts, claims, and reasoning steps, each with distinct levels of uncertainty.
    %
    To address this gap, this study investigates uncertainty granularity (\ie, the extent to which uncertainty is expressed at different levels within an LLM response) and examines its impact on LLM-assisted decision-making.
    We conducted a large-scale, between-subjects study (N=192) in which participants answered challenging medical questions using LLMs that displayed uncertainty at three different granularities: output-level (entire response), relation-level (individual reasoning steps), and token-level (specific words). 
   Our findings reveal distinct behavioral effects as a function of uncertainty granularity. Token-level uncertainty increased users’ agreement with the AI, whereas output- and relation-level uncertainty did not increase agreement but instead reduced users’ confidence in their own answers. Notably, relation-level uncertainty also reduced external verification (\ie, internet searches, checking provided URLs), steering users away from independent fact-checking and toward reliance on the LLM and its accompanying uncertainty cues.
    Our findings demonstrate that uncertainty granularity significantly shapes how users interact with and verify LLM outputs, providing concrete design guidance for building responsible LLM applications that encourage appropriate skepticism and verification behaviors.

  
\end{abstract}



\maketitle
\section{Introduction}

Despite their remarkable capability and widespread adoption, Large Language Models (LLMs) can produce misleading errors and require further verification from users.
At the same time, a growing body of research has suggests that users often exhibit uncalibrated trust in AI systems, over-relying on incorrect outputs without sufficient critical reflection~\cite{bansal2021does, zhang2020effect, buccinca2021trust, lu2021human, steyvers2025large}.
Therefore, calibrating user trust in LLMs is crucial for ensuring effective and responsible use of AI, particularly in high-stakes domains such as finance, law, and healthcare.


One widely advocated approach for improving trust calibration is communicating AI's uncertainty of their own outputs.
Researchers have investigated both technical methods for uncertainty quantification (UQ) and behavioral impacts of AI uncertainty on human-AI teaming.
On the one hand, showing AI uncertainty can prompt users to slow down, engage in more analytical thinking, reduce over-reliance, and ultimately improve team performance~\cite{prabhudesai2023understanding, zhang2020effect, rezaeian2025explainability}.
On the other hand, the effectiveness of uncertainty on trust calibration is influenced by various factors such as user background~\cite{cao2024designing, rezaeian2025explainability}, data and task~\cite{cau2023effects, zhao2023evaluating}, and the format of uncertainty presentation (\eg, visualization, numbers)~\cite{cao2024designing, zhao2023evaluating, lee2025towards}, highlighting the importance of carefully designing and evaluating uncertainty communication.

\begin{figure}
  \centering
  \includegraphics[width=0.95\linewidth]{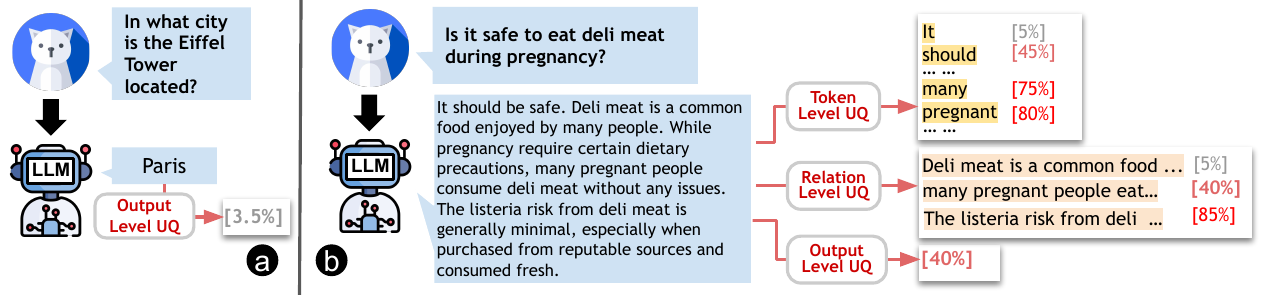}
  \vspace{-1em}
  \caption{\textbf{Uncertainty granularity in LLM responses}. (a) Output-level uncertainty displays a single score for the entire response. (b) Finer-grained uncertainty highlights varying uncertainty across reasoning steps (relation-level) and individual words (token-level). Cat and AI assistant icons by Freepik from Flaticon (\url{https://www.flaticon.com})}
  \Description{Two side-by-side examples of an LLM answering a medical question. Panel (a) shows output-level uncertainty as a single confidence bar beneath the entire response. Panel (b) shows the same response with finer-grained uncertainty: a tree-like reasoning structure with separate uncertainty scores per reasoning step, and individual words highlighted in varying shades of red according to per-word uncertainty. }
  \label{fig:teaser}
\end{figure}

However, these insights are primarily derived from AI on classification tasks, where uncertainty is communicated at output-level as probability scores.
As shown in \autoref{fig:teaser}, the generative nature of LLMs introduces a fundamental shift regarding the granularity of uncertainty.
In contrast to discrete labels on classification, LLM outputs are often free-form text that can contain multiple claims and reasoning steps, each of which potentially carries different levels of uncertainty. 
For example, when answering a medical question (\autoref{fig:teaser}b), an LLM might be highly confident about general facts, but uncertain about specific information on listeria, all within the same output. 
This shift opens a new design space for both quantifying and communicating LLM uncertainty.
Recent advances in LLM UQ can now estimate uncertainty at finer granularities, such as highlighting specific words where LLM hesitates~\cite{duan2024shifting, fadeeva2024fact}, or quantifying uncertainty separately for different reasoning paths and intermediate steps~\cite{freedman2025argumentative, da2025understanding, yin2024reasoning}. 
Despite these technical advances in LLM UQ, our understanding of how LLM uncertainty impacts user trust and task performance remains limited.

Uncertainty granularity is a critical design factor for two key reasons. First, extensive research has established that providing the right amount of information (\ie, neither too little nor too much) is essential for effective human-AI collaboration~\cite{Guesmi2023InteractiveEWA, kizilcec2016much}. Second, granularity fundamentally shapes how users localize the source of uncertainty, which in turn affects their perception and behaviors when using LLMs.
A few recent studies have begun exploring LLM uncertainty communication, but have focused primarily on a single granularity. For example, \citet{metzger2024empowering} found that over-trust can be reduced by adding natural language disclaimers about LLM limitations.
\citet{vasconcelos2025generation} investigated token-level uncertainty impact user performance in AI code completion. 
As a result, the user impacts of LLM uncertainty granularity remain poorly understood.

This study aims to fill this critical gap by examining how the \textit{uncertainty granularity} affects users in LLM-assisted decision-making.
Drawing on recent advances in LLM UQ~\cite{shorinwa2025survey}, we identify three common granularity levels: token-level (uncertainty for individual words), relation-level (uncertainty for logical steps), and output-level (uncertainty for the entire response).
We then compare the three uncertainty granularities with a baseline that provides no uncertainty information in helping participants answer challenging medical questions, via a large-scale between-subjects study ($N = 192$).
Our results reveal that \textit{more detailed uncertainty information does not necessarily yield more cautious use}. 
\added{In particular, the finest-grained presentation (token-level) increased agreement with the LLM (\ie, participants answer is the same as LLMs), while the intermediate presentation (relation-level) increased reliance on the LLM's uncertainty cues (\ie, reduced independent fact-checking via internet search or URL links), without a corresponding increase in agreement.} 
Across conditions, these behavioral shifts did not translate into reliable gains in task accuracy, which may be attributable to the inherent difficulty of the medical questions.
Together, these findings demonstrate that uncertainty granularity meaningfully shapes user reliance and verification behaviors, positioning it as a critical design factor for responsible LLM systems.

%

\section{Related Work}

\subsection{Quantifying LLM Uncertainty}

Based on their mechanisms, LLM UQ methods can be categorized into three main groups: generation probability, self-verbalization, and semantic-similarity~\cite{shorinwa2025survey}. 
Generation probability methods utilize the probability distribution over token generation to calculate UQ metrics, such as entropy and average token log-probability~\cite{manakul2023selfcheckgpt, bakman2024mars, malinin2020uncertainty}.
Self-verbalization methods prompt LLMs, often combined with confidence calibration, to report uncertainty alongside their outputs~\cite{xiong2023can, kadavath2022language, band2024linguistic}. 
Semantic similarity methods quantify LLM uncertainty based on consistency across multiple responses to the same or similar inputs~\cite{kossen2024semantic, qiu2024semantic, kuhn2023semantic}.
Importantly, even within the same UQ approach, uncertainty can be communicated in a variety of formats.
For instance, generation probability can be shown at the token level~\cite{fadeeva2024fact}, aggregated into reasoning steps~\cite{yin2024reasoning}, or for the entire response~\cite{bakman2024mars}. 
Similarly, self-verbalization can express uncertainty as scores for individual reasoning steps~\cite{tanneru2024quantifying}, or complete outputs~\cite{tao2024trust}, or through natural language expressions~\cite{band2024linguistic}.
However, prior work on LLM UQ has primarily focused on improving the faithfulness of these UQ methods, providing limited discussion into how different UQ communication influence users in perceiving and using LLMs.
At the same time, a few recent empirical studies have shown that uncertainty communication critically shapes human reliance and behavior. 
For example, \citep{kim2024m} showed that uncertainty expressed as \textit{``I'm not sure, but...''} reduce user reliance when models are wrong, but is less effective when phrased as \textit{``there is an uncertainty, but...''}. 
This study addresses this gap by investigating uncertainty granularity, a critical yet underexplored factor in uncertainty communication that influence how users make decisions.

\subsection{Measuring and Calibrating Trust in Human-AI Collaboration}
As an important factor for human-AI collaboration, trust has received substantial research attention, yielding diverse measurement and calibration approaches. 
Below, we review the literature most germane to our study.

One widely-used setting for investigating trust is the judge-advisor paradigm \cite{buccinca2021trust, bansal2021does, kim2024m}.
In this paradigm, participants (judges) receive recommendations from AI (advisor) and submit their own final decisions. 
Trust is operationalized through complementary behavioral and self-report measures. 
Behavioral measures include agreement rate between participants and AI, and the accuracy of participants final answers.
Self-report measures commonly assess participants’ confidence in their own judgments and their perceived trust in AI system. 

Given these measurement frameworks, researchers have examined diverse interventions designed to improve trust calibration.
These interventions can be categorized into three main mechanisms: information provision (\eg, uncertainty, explanations) \cite{bansal2021does, wang2022extending, yang2020visual}, cognitive engagement (\eg, cognitive forcing functions, Socratic questioning) \cite{buccinca2021trust,danry2023don}, and user skill development (\eg, onboarding about AI limitations) \cite{cai2019hello, lee2024interactive}. 
Despite of some success, trust calibration remains challenging. 
The effectiveness of trust calibration is influenced by a variety of factors and is not always guaranteed.
For example, explanation-based approaches have been found to be ineffective or even increase overreliance~\cite{buccinca2021trust, bansal2021does}. 

Recent work has begun to identify specific design factors that moderate intervention effectiveness. Within information provision approaches, which is the focus of our study, what to present and how to present has emerged as critical determinants of outcomes. 
\added{Regarding what to present, prior work has identified a central tension between informativeness and cognitive overload~\cite{Guesmi2023InteractiveEWA, kizilcec2016much}.
For example, \citet{he2025finegrained} found that exposing the multi-step LLM decision workflow to users help them develop critical mindset, but also increased cognitive overload and did not improve team performance.
\citet{salimzadeh2024dealing} found that the whether enough information has been provided to enable accurate decision (diagnostic vs. prognostic tasks) shaped trust in human--AI collaboration.
}
\added{Regarding how to present,}
\citet{yang2020visual} and \citet{wang2022extending} found that certain visualization formats yield better trust calibration and team performance than others for visual explanations. 
The importance of format extends to uncertainty communication as well \cite{cao2024designing, lee2025towards, zhao2023evaluating, xu2025confronting}.
For example, 
\citet{lee2025towards} showed that visualizing uncertainty as distances in embedding space can improved task performance.
\citet{zhao2023evaluating} demonstrated that ordinal uncertainty visualizations can more effectively calibrate user behavior in model usage.


Despite these advances, existing studies provide little discussion about the granularity of uncertainty, an important design parameter given recent advances in LLM UQ. This study aims to address this gap. 




\section{Uncertainty Quantification and Visualization}

\subsection{Design Rationales}
\label{subsec:design-rationales}

\begin{figure}
    \centering
    \includegraphics[width=\linewidth]{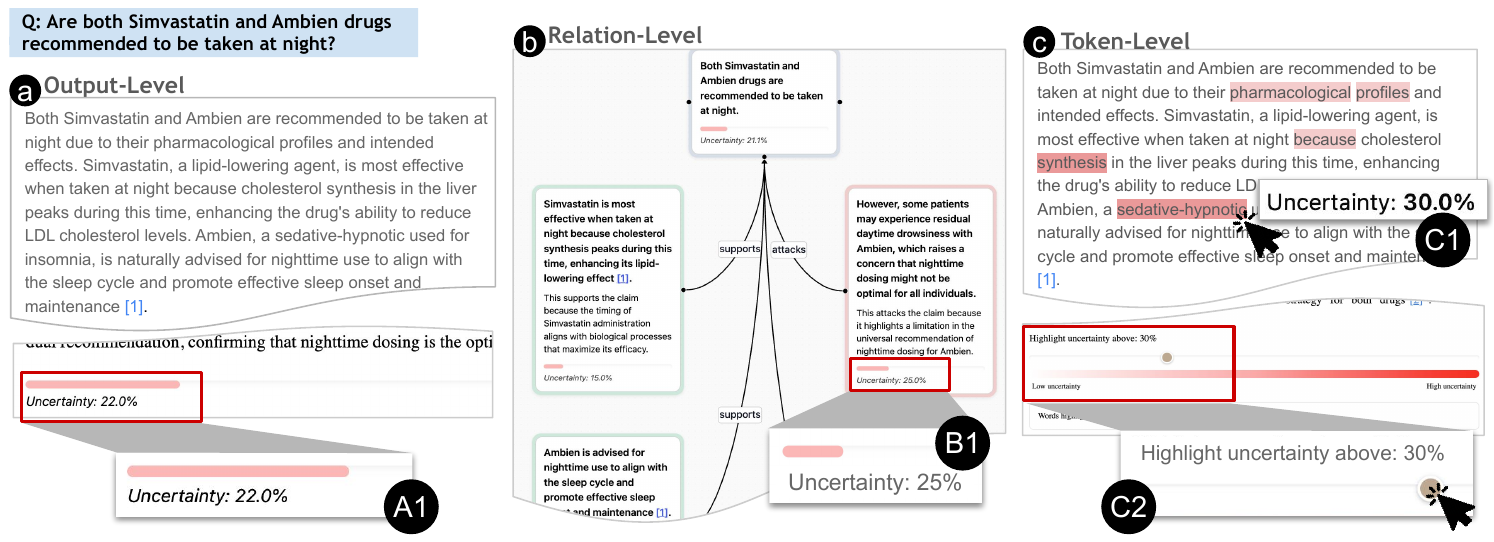}
    \caption{LLM uncertainty quantification and visualization across \outputUQ{} (a), \relationUQ{} (b), and \tokenUQ{}(c).}
    \Description{Three panels illustrating the three uncertainty visualization conditions used in the study. Panel (a) shows output-level UQ as a horizontal red progress bar with a percentage beneath the LLM response. Panel (b) shows relation-level UQ as a node-link diagram with a central claim node connected to supporting and attacking sub-argument nodes, each labeled with their own uncertainty score. Panel (c) shows token-level UQ as the response text with individual words highlighted in red at varying intensities, plus a slider for adjusting the highlight threshold.}
    \label{fig:UQ-formts}
\end{figure}

Providing the appropriate amount of information is a central design challenge in human--AI systems, requiring a balance between being informative and avoiding cognitive overload~\cite{Guesmi2023InteractiveEWA, kizilcec2016much}. We argue that a similar trade-off arises when communicating uncertainty in LLM outputs. While coarse uncertainty information may be easy to interpret, it can obscure important nuances. In contrast, fine-grained uncertainty information may offer richer signals but risk overwhelming users.
Motivated by this tension, we design and develop interfaces (\autoref{fig:UQ-formts}) to investigate how different levels of LLM uncertainty granularity (\ie, at which detailed level uncertainty is expressed within an LLM response) affects human--LLM interaction. 
Specifically, we pose two research questions. 
\begin{itemize}
    \item \textbf{RQ1:} How does LLM \textit{uncertainty granularity} influence user perception and behavior in decision-making? 
    \item \textbf{RQ2:} How does uncertainty granularity interact with \textit{contextual factors}, such as AI correctness, to influence user perception and behavior?
\end{itemize}


\added{
Note that differences in UQ methods and visualizations across uncertainty granularity conditions reflect real-world constraints rather than a design oversight: each granularity level corresponds to fundamentally different UQ methods requiring distinct visualizations. Imposing a uniform approach would produce ecologically invalid stimuli and introduce bias by forcing some conditions to use suboptimal visualizations.
}

\noindent
\textbf{Controlling for Response Content.} 
While different UQ methods may produce different response content in practice due to their different uncertainty calibration mechanism, we enforce same response content to isolate the effects of uncertainty granularity, the primary focus of this work.
Specifically, we generate all responses using \texttt{gpt-4o-mini} with chain-of-thought prompting to include reasoning steps and explanations, which has been shown to improve both response quality and uncertainty estimation~\cite{podolak2025read, becker2024cycles}.

\noindent
\textbf{Choice of UQ Methods.} There are numerous algorithms exist for estimating uncertainty at each granularity level. 
However benchmarking their faithfulness is a challenging research question beyond the scope of this study. Instead, we selected commonly used and accessible methods that best serve our experimental goal: uncertainty at each granularity that reflects the current LLM UQ practice while maintaining consistency in response content. 
Specifically,
for token-level UQ, we use the generation probabilities of each token provided by the LLM.
For output-level, we directly prompt the LLM to attach an numerical score along with their response.
For the relation-level, we prompt LLM to divide the responses into reasoning steps and then articulate the uncertainty for each.

\noindent
\textbf{Manual Reviewing.}
For the LLM responses used in the user study (\autoref{sec:study}), all authors manually reviewed the generated responses and uncertainty information to ensure consistency across levels of granularity, avoiding obvious conflicts (\eg, low output-level uncertainty paired with uniformly high relation-level uncertainties).
\added{Specifically, one author compiled all generated responses and their corresponding uncertainty scores, visualizing them in the same interface used in the user study while enabling side-by-side comparison across the three UQ conditions. The remaining two authors independently reviewed each question, flagging cases where the perceived uncertainty magnitude were substantially misaligned across conditions. 
Any flagged cases were then discussed among all three authors until consensus was reached.}
Importantly, responses were not selected based on the quality of the UQ.
This decision was made to realistically reflect the behavior of current LLM UQ methods, which often remain overconfident even when producing incorrect answers \cite{shorinwa2025survey}.

In the remainder of this section, we describe how we generate and visualize uncertainty information at each granularity, as shown in the \autoref{fig:UQ-formts}.



\subsection{Output-Level}
In the output-level uncertainty condition, uncertainty is communicated as a single confidence score for the entire response (\autoref{fig:UQ-formts}-a). 
Output-level uncertainty is the standard method for comparing and ranking various LLMs \cite{shorinwa2025survey}.
We generate this score through self-verbalization, prompting the LLM to provide a numerical estimate in the range $[0\%,100\%]$ alongside its response. 
While LLMs are known to exhibit overconfidence, recent work has shown that incorporating chain-of-thought reasoning can improve the faithfulness of self-reported uncertainty estimates~\cite{podolak2025read}.

%

As shown in \autoref{fig:UQ-formts}-A1,
we visualize output-level uncertainty as a horizontal progress bar, where the length of the red segment encodes the uncertainty magnitude and the gray background represents the full scale (100\%).

\subsection{Relation-Level}
Relation-level uncertainty is calculated for each reasoning components, operating at a finer granularity than output-level but coarser than token-level (\autoref{fig:UQ-formts}b). 
This condition reflects an important direction in LLM UQ research motivated by two key observations. First, modern LLMs increasingly employ multi-step reasoning processes, where different reasoning steps may carry substantially different levels of uncertainty~\cite{yin2024reasoning}. Second, understanding where uncertainty arises in the reasoning chain is critical to identify and mitigate potential errors~\cite{freedman2025argumentative, da2025understanding}.

While numerous methods exist for quantifying uncertainty at the reasoning steps~\cite{yin2024reasoning, freedman2025argumentative, da2025understanding}, we employ self-verbalization rather than aggregating token generation probabilities or calculating semantic similarity. Our pilot studies indicated that self-verbalized uncertainty are more interpretable for users.
Specifically, we adopt the framework proposed by \citet{freedman2025argumentative}. 
For each question and the generated response, the response is divided into a \textit{central claim} and a list of \textit{sub-arguments}. Each \textit{sub-arguments} are labeled either \textsc{SUPPORTS} or \textsc{ATTACKS} regarding their relation to the \textit{central claim}. 
The LLM is prompted to generate numerical uncertainty score in the range $[0\%, 100\%]$ for the central claim and for each sub-argument. We then adopt the method in \cite{freedman2025argumentative} to calibrate the self-verbalized uncertainty. 

The resulting argumentative structure is presented to users as an interactive node-link visualization. The central claim is shown as a primary node with its aggregated uncertainty, while supporting and attacking sub-arguments are displayed as connected nodes with their individual uncertainty values. This representation allows users to inspect uncertainty at the level of reasoning steps and to observe how different sub-arguments contribute to the overall uncertainty of the model’s answer.

\subsection{Token-Level}

In the token-level uncertainty condition, uncertainty is communicated at the granularity of individual words, as illustrated in \autoref{fig:UQ-formts}-c.
During response generation, the LLM outputs a log probability (\texttt{logprob}) for each token, representing the logarithm of the probability of generating that token.
To improve the readability for users,
we first remove non-semantic tokens (\eg, punctuation and formatting symbols) and then aggregate subword tokens belonging to the same word by averaging their log probabilities. For example, the word ``uncertainty'' may be tokenized into subwords like \texttt{[un, certain, ty]}. We average the log probabilities of these three tokens to obtain a length-normalized uncertainty estimate for the complete word.
We then convert the log probability into uncertainty scores for each word using $(1 - exp(logprob))$.



We follow the common practice to visualize token-granularity uncertainty as heatmap. 
The uncertainty of individual tokens is encoded using the color intensity of their text background. 
Darker red indicating higher uncertainty. Users can hover over any individual word to view its precise uncertainty value (\autoref{fig:UQ-formts}-C1). 
To mitigate potential information overload caused by dense color encoding, we employ an interactive highlighting mechanism. As shown in \autoref{fig:UQ-formts}-C2, users can adjust a threshold and only tokens whose uncertainty exceeds the threshold will be visually emphasized.


\section{Study Design}
\label{sec:study}

\subsection{Tasks, Procedure, and Experimental Conditions}

We conducted a between-subjects experiment with within-subject comparisons at the trial level. Participants were randomly assigned to one of four conditions: 
(1) \baseline{} (no uncertainty), 
(2) \tokenUQ{} (uncertainty for words), 
(3) \relationUQ{} (uncertainty for reasoning components), or 
(4) \outputUQ (uncertainty for the entire response). Each participant completed a set of information-seeking questions. 

All questions are challenging (\ie, beyond what most laypeople or popular search engines can readily answer) and factual (\ie, objectively verifiable) questions adapted from the MedQuAD \cite{ben2019question} dataset, following the practices of~\citet{kim2024m}. We provided fixed responses with uncertainty generated in advance rather than allowing users to interact with an LLM system in real time. This approach ensures that the LLM responses realistically reflect real-world LLM behavior while maintaining consistency across all participants. 
The order of the eight questions was randomized, with balanced coverage of AI correctness (four correct and four incorrect responses) and uncertainty magnitude.
Additional details on response generation are provided in \autoref{subsec:design-rationales}.

\begin{figure}
    \centering
    \includegraphics[width=\linewidth]{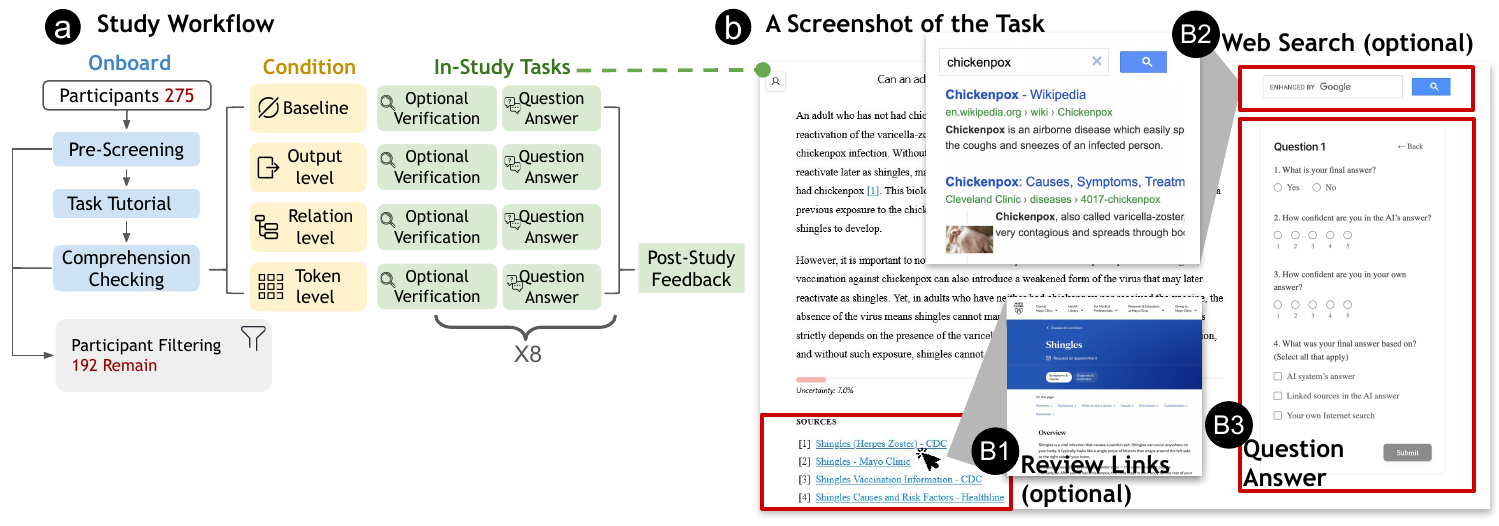}
    \caption{\textbf{Study Procedure and Task.} (a) An overview of the user study workflow. Participants completed onboarding, answered eight information-seeking questions under one of four conditions, and then completed a post-study questionnaire.
    (b) A screenshot of the in-study task interface. Participants are required to complete the question-answering in-study task (B3) for all eight questions under the assigned condition. They may optionally verify their answers by reviewing provided links (B1) and using the web search tool (B2).}
    \Description{Two-panel figure. Panel (a) is a horizontal flowchart of the study procedure with three stages: onboarding (consent, demographics, tutorial), main task (eight question trials with one of four UQ conditions), and post-study questionnaire. Panel (b) is an annotated screenshot of the task interface, showing the medical question and LLM response on the left, a panel of clickable reference links labeled B1, an embedded Google Search panel labeled B2, and the answer-input area with confidence ratings labeled B3.}
    \label{fig:workflow}
\end{figure}

As shown in \autoref{fig:workflow}-a, the study consisted of three phases. First, participants completed a background questionnaire assessing demographics and prior experience with LLMs, followed by a task tutorial and comprehension questions.
Second, participants completed eight information-seeking questions. For each question, they were presented with an LLM-generated response corresponding to their assigned condition. 
\autoref{fig:workflow}-b shows a screenshot of the in-study task interface.
Before making the binary decision, participants could consult URL links provided in the LLM response or use an embedded Google Search panel. 
Participants reported their confidence in both the LLM’s response and their own answer on 5-point Likert scales. 
Participants also indicated, via a multiple-choice question, whether and which verification actions informed their decisions.
Finally, participants completed a post-study questionnaire assessing their trust, confidence, and perceptions of the AI system using 5-point Likert-scale items, and were given the option to provide free-form feedback about (i) their experience of using the interface and (ii) their motivation for (not) using the verification tools.
More details about the questions used in the in-study task and post-study questionnaires can be found in \autoref{sup:questions}.


\subsection{Dependent Variables}

For our dependent variables (DVs), we adapt the measures used by \citet{kim2024m}, who examined the impact of LLM uncertainty on user reliance and trust.
In addition to these measures, we also analyzed interaction logs to track participants' verification behaviors—specifically, whether they follow LLM-provided links or conduct independent internet searches when answering questions. This enables us to examine how different uncertainty granularity influence not only users' subjective responses but also their actual verification behavior.

We analyze both trial-level and participant-level dependent variables, including objective behavioral measures and subjective self-reports. Our dependent variables are grouped as follows.

\begin{itemize}[leftmargin=*]
    \item \textbf{Objective, trial-level measures:}
    \begin{itemize}
        \item \textit{Agreement}: Whether participants accept the LLM's response as their final answer. 
        \item \textit{Time}: Time taken to answer each question.
        \item \textit{Accuracy}: Correctness of participants' final answers.
        \item \textit{Verification behavior}: Whether, how often, and the type of 
        verification participants engage. 
        We collected objective measures from 
        interaction logs of participants, as well as subjective measures where 
        participants reported the information resources (internet search, URL links, or LLM responses) they used.
    \end{itemize}

    \item \textbf{Subjective, trial-level measures:}
    \begin{itemize}
        \item \textit{Confidence in AI}: Participants reported confidence in the LLM's responses on a 5-point scale.
        \item \textit{Confidence in own answers}: Participants reported confidence in their final answers on a 5-point scale.
        \item \textit{Resources used}: Participants specified the information sources they relied on when answering the questions, including the LLM response, provided links, and self-initiated internet searches.
    \end{itemize}

    \item \textbf{Subjective, participant-level measures:}
    \begin{itemize}
        \item \textit{Perception of LLM systems}: Post-study measures of trust and perceived transparency about the LLM systems on a 5-point scale, adapted from \citet{kim2024m}.
    \end{itemize}
\end{itemize}

\subsection{Hypothesis and Analysis}
\label{subsec:hypo}
We test following hypotheses regarding the effects of uncertainty granularity on dependent variables (DVs).

{\bf Effects of uncertainty granularity (RQ1).} 
First, we hypothesize that condition (\ie, UQ granularity as output level, relation level, and token level) affects participant behavior and perceptions. We test this hypothesis using a between-condition analysis. For trial-level DVs, we fit mixed-effects models of the form
\texttt{DV $\sim$ Condition + (1|participant) + (1|question)}, with \texttt{Control} as the reference level for \texttt{Condition}. After fitting the model, we use Wald tests to assess whether the estimated effects of different UQ conditions differ significantly from one another. 
For participant-level DVs, we conduct a one-way analysis of variance (ANOVA) to compare means across conditions. When the omnibus test is significant, we perform post-hoc pairwise comparisons using Tukey’s test.
%

{\bf Effects of AI correctness and its interaction with uncertainty granularity (RQ2).}
Prior findings has suggested that contextual factors such as AI correctness may also influence user behavior. We hypothesize that uncertainty granularity modulates the influence of these contextual factors.
To examine this, we conduct within-condition analyses restricted to conditions that include uncertainty expression. For each such condition, we fit mixed-effects models of the form
\texttt{DV $\sim$ \texttt{AICorrectness} * Condition + (1|participant) + (1|question)}, where \texttt{AICorrectness} indicate whether the AI response is correct or not.
\added{For all mixed-effects models, we include random intercepts for participants and questions, but not random slopes. Given the binary outcomes and only eight trials per participant, random-slope models are prone to convergence issues and are difficult to estimate reliably. Random-intercept models are common in HCI empirical studies~\cite{matuschek2017balancing} and appropriately account for repeated observations within participants.
}

\subsection{Participants and Data Collection}

We conducted the study on Prolific, a widely-used crowd-sourcing platform for human subject experiments. 
Participants were screened to ensure U.S. residency in accordance with our approved IRB protocol and a Prolific approval rating of at least 95\% to ensure data quality. 
Additional criteria included basic familiarity with data analysis, an undergraduate degree or higher, prior experience with LLMs, and no medical-related occupation.
These additional criteria are used to ensure that participants could meaningfully engage with AI-generated medical information and uncertainty cues, rather than answering the questions merely based on their own knowledge.

\added{
We conducted power analysis to determine the minimum required sample size  ($ICC = 0.20, \alpha = 0.05, power = 0.80$). 
Assuming a medium effect size based on pilot data ($Cohen's f = 0.20$ for continuous outcomes; $OR = 2.0$ for binary outcomes), the analysis indicates a minimum of 45 participants per condition (180 total). 
Based on our prior experience on Prolific samples, we anticipated a high exclusion rate ($\sim$33\%) and thus targeted 270 participants. 
We recruited 275 in total, of whom $83$ were excluded for failing screening or attention checks, yielding a final analytic sample of 192 participants (48 per condition).
}

The study took an average of $23.8$ minutes to complete. Participants were compensated at a rate of \$12 per hour. 
The participant pool exhibited diverse coverage in age and LLM experience, including variation in usage frequency and application contexts.
\added{Participants were randomly assigned to four experimental conditions, with no significant differences in demographics or LLM experience across groups.}
Detailed demographics are reported in \autoref{sup:participants}.

\section{Confirmatory Analysis Results}
\label{sec:confirm-results}
Here, we report the results of our confirmatory analyses regarding the hypothesis in \autoref{subsec:hypo}. We present estimated means for continuous DVs, and estimated probabilities for binary DVs, both are denoted as $M$. In all cases, values are derived from the fitted models and accompanied by 95\% confidence intervals ($95\%CI$), computed without conditioning on the random effects. 
We use the term {\it significant} to denote statistical significance at the $p < .05$ level. Results are reported in \autoref{fig:result-1} and \autoref{tab:result-2}. 
\subsection{Agreement with AI}
\added{Agreement with the AI is measured by whether participants’ final answers match the AI’s answer. 
This is a commonly used behavioral measure in human-AI collaboration.}

\subsubsection{Direct effects of uncertainty granularity (RQ1).}
\begin{figure}
    \centering
    \includegraphics[width=0.8\linewidth]{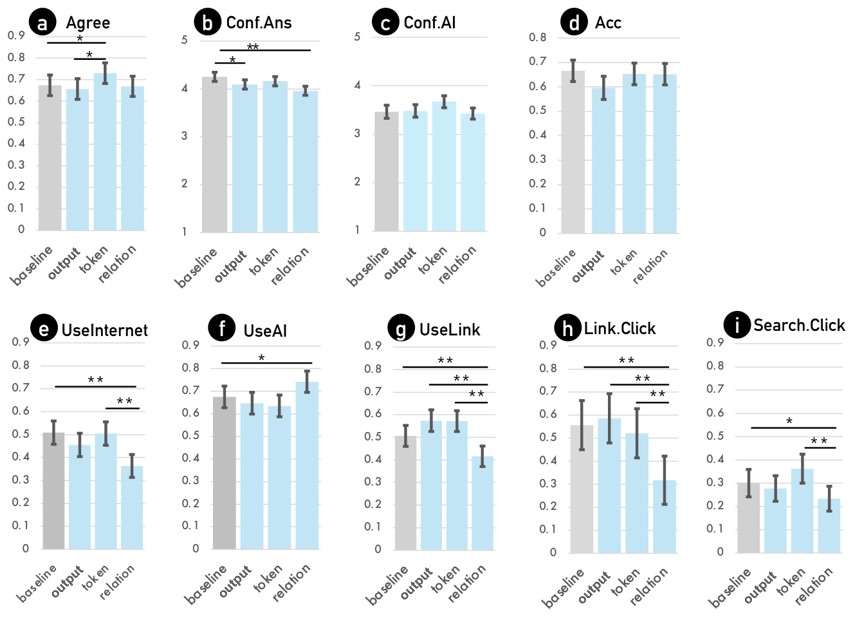}
    \vspace{-1em}
    \caption{Comparison of DVs across conditions. We show model-estimated marginal means from the confirmatory mixed-effects analyses and error bars indicate 95\% confidence interval. * denoting $p\le.05$ and ** denoting $p\le.01$. For reference, higher values are generally desirable for Acc, UseLink, UseInternet, LinkClick, and SearchClick, reflecting greater accuracy and verification behavior. For Agree and confidence measures (Conf.Ans, Conf.AI), desirability depends on context: higher agreement is appropriate when the AI is correct but reflects overreliance when incorrect, and higher confidence is desirable only when well-calibrated. }
    \Description{A grid of bar charts comparing the four study conditions (Baseline, UQ-Output, UQ-Relation, UQ-Token) across multiple dependent variables: agreement with AI, confidence in own answer, confidence in AI, accuracy, self-reported use of links, self-reported use of internet, link clicks, and web searches. Each bar shows the model-estimated mean with 95\% confidence interval error bars, and significant differences between conditions are marked with asterisks.}
    \label{fig:result-1}
\end{figure}

Participants' agreement with AI varied significantly across conditions (\autoref{fig:result-1}-a). 
\textbf{The \tokenUQ{} condition elicited significantly higher agreement with AI} ($M = 0.730$, 95\% CI $[0.682, 0.778]$) compared to the baseline ($M = 0.674$, 95\% CI $[0.626, 0.722]$, $p \le .05$, \added{OR = 1.62}). In contrast, neither \outputUQ{} ($M = 0.657$, 95\% CI $[0.609, 0.705]$) nor \relationUQ{} ($M = 0.669$, 95\% CI $[0.622, 0.716]$) conditions differed significantly from baseline.


\subsubsection{Modulating effect of uncertainty granularity (RQ2).}


Across all conditions, participants showed significant higher agreement when the LLM was correct than when it was incorrect (all $p < .01$). In the \baseline{} condition, agreement increased from 0.517 for incorrect responses to 0.827 for correct responses. Similar patterns emerged in the uncertainty conditions: \outputUQ{} (0.551 to 0.766), \tokenUQ{} (0.573 to 0.886), and \relationUQ{} (0.519 to 0.820).
We found no significant AI Correctness~$\times$~Condition interaction for agreement, \ie, \textbf{\added{the effect of AI correctness on agreement does not change significantly across conditions}}.

\begin{table}[t]
\centering
\caption{Within-condition analysis. For each condition, we compare DVs measured on trials with correct and incorrect LLM responses. * denoting $p\le.05$,  ** denoting $p\le.01$, and *** denoting $p\le.001$.}
\Description{Within-condition analysis. For each condition, we compare DVs measured on trials with correct and incorrect LLM responses. * denoting $p\le.05$,  ** denoting $p\le.01$, and *** denoting $p\le.001$.}
\label{tab:result-2}
\vspace{-1em}
\includegraphics[width=\linewidth]{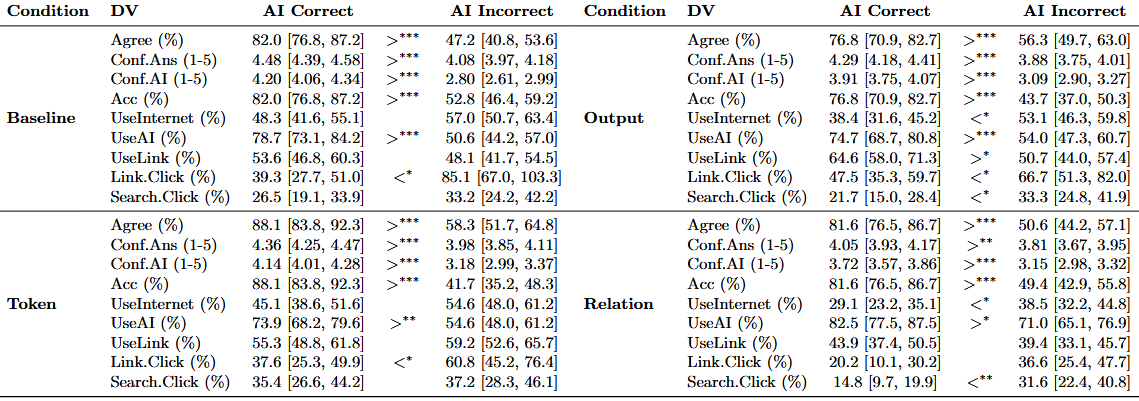}
\end{table}

\subsection{User Verification}
For each question, participants could optionally verify the LLM’s response by examining the provided links or by conducting their own internet search. 
Participants also reported whether they relied on the AI, the provided links, or their own internet search when answering the question.
\added{
We used objective verification behaviors and self-reported information sources as two primary indicators of whether participants relied on the LLM's uncertainty cues.
}

\subsubsection{Direct effects of uncertainty granularity (RQ1).}
As shown in \autoref{fig:result-1}, uncertainty granularity significantly shaped users’ verification behavior, as reflected in both self-reported strategies and interaction logs.

Based on participants’ self-reports, \textbf{participants in the \relationUQ{} condition reported higher reliance on AI} ($M = 0.743$, 95\% CI $[0.696, 0.790]$) compared to the \baseline{} condition ($M = 0.675$, 95\% CI $[0.627, 0.723]$, $p < .05$, \added{OR = 1.87}). Neither \tokenUQ{} ($M = 0.635$, 95\% CI $[0.587, 0.683]$) nor \outputUQ{} ($M = 0.647$, 95\% CI $[0.599, 0.696]$) conditions differed significantly from baseline. Post-hoc pairwise tests revealed that \outputUQ{} participants reported using links more frequently than \relationUQ{} participants ($p < .05$, \added{OR = 2.02}). Qualitative feedback corroborated this pattern, with participants in \relationUQ{} noting reduced verification motivation: \textit{``I did not [verify] because I found the AI tool to summarize all key information''}.

Behavioral measures confirmed these self-reported patterns. 
Here, link click and web search are both treated as binary trial-level DV.
Compared to baseline, \textbf{\relationUQ{} participants were less likely to click links} ($M = 0.318$, 95\% CI $[0.213, 0.422]$ vs. $M = 0.557$, 95\% CI $[0.450, 0.664]$, $p < .05$, \added{OR = 0.42}) \textbf{and conducted web searches} ($M = 0.247$, 95\% CI $[0.184, 0.310]$ vs. $M = 0.342$, 95\% CI $[0.278, 0.406]$, $p < .05$, \added{OR = 0.81}). No significant reductions in link clicking or web search usage were observed for \outputUQ{} or \tokenUQ{} conditions relative to baseline.

Interestingly, \textbf{the condition with higher AI agreement (\tokenUQ{}) does not necessarily correspond to lower verification behavior (\relationUQ{})}. 
One possible explanation is that external verification is reduced in the \relationUQ{} condition because the system presents relational uncertainty information in a clear and effective manner, allowing participants to assess reliability internally rather than seeking additional external sources.
This contrast highlights the importance of how uncertainty is communicated, rather than simply whether it is provided, in shaping user trust and engagement.

\subsubsection{Modulating effect of uncertainty granularity (RQ2).}

\textbf{Compared with when the LLM was incorrect, participants showed reduced external verification behavior when the LLM was correct in \outputUQ{} and \relationUQ{}.}
We observed a significant AI Correctness~$\times$~Condition interaction for link usage ($p < .05$). In the \outputUQ{} condition, participants were less likely to report using internet search when the AI was correct ($M=0.397, 
95\%CI=[0.324,0.469]$) compared to when it was incorrect ($M=0.512, 
95\%CI=[0.441,0.583]$) ($p < .05$, \added{OR = 0.55}).
In the \relationUQ{} condition, participants were less likely to click links when the AI was correct ($M=0.118, 95\%CI=[0.067,0.170]$) compared to when it was incorrect ($M=0.204, 95\%CI=[0.152,0.256]$) ($p < .05$, \added{OR = 0.49}), and performed fewer web searches when the AI was correct ($M=0.157, 95\%CI=[0.072,0.241]$) compared to when it was incorrect ($M=0.339, 95\%CI=[0.255,0.423]$) ($p < .05$, \added{OR = 0.43}). No significant correctness-related differences in verification behavior in the \baseline{} or \tokenUQ{} conditions.



\subsection{Self-reported Confidence}
For each question, participants reported both their confidence in the LLM’s response and their confidence in their submitted final answer, using a 5-point Likert scale.

\subsubsection{Direct effects of uncertainty granularity (RQ1).}
As shown in \autoref{fig:result-1}-c, we observed \textbf{no significant difference across the four conditions in participants’ confidence in the LLM’s responses} (\baseline{}: $M=3.578, 95\%CI=[3.432,3.724]$; \tokenUQ{}: $M=3.615, 95\%=[3.471,3.760]$; \relationUQ{}: $M=3.442, 95\%=[3.300,3.583]$; \outputUQ: $M=3.498, 95\%=[3.354,3.642]$). 
We speculate that this may be because participants reported their confidence in LLM responses based on the provided uncertainty information, which was largely consistent across conditions.

However, the granularity of uncertainty significantly influences user confidence in their own answers (\autoref{fig:result-1}-b).
{\bf Compared with the \baseline{} ($M=4.253, 95\%CI=[4.156,4.349]$), both \outputUQ{} ($M=4.094, 95\%CI=[3.998,4.190]$, \added{$d=-0.23$}) and \relationUQ{} ($M=3.962, 95\%CI=[3.867,4.056]$, \added{$d=-0.40$}) conditions significantly decreased user confidence in their own answers , whereas the Token-Level condition did not ($M=4.161, 95\%CI=[4.065,4.258]$).}
Via post-hoc Wald test, participants in \tokenUQ{} reported higher confidence in their own answers than those in \relationUQ{} ($p < .05$, \added{$d = -0.34$}). 
Combined with the results that \tokenUQ{} condition led to significantly higher AI agreement, this applies that participants in \tokenUQ{} were more likely to agree with AI and in a more confident manner.

\subsubsection{Modulating effect of uncertainty granularity (RQ2).}

Across all conditions, participants reported significantly higher confidence when the LLM was correct than when it was incorrect, both in the LLM’s response and in their own final answers (all $p<.05$). This correctness-related increase in confidence was consistent across conditions, and we found \textbf{\added{no significant AI Correctness~$\times$~Condition interaction}} for user confidence.

\subsection{DVs Without Significant Effects}
Contrary to our hypotheses, we did not observe significant differences across conditions in task accuracy, task completion time, or participants’ self-reported perceptions of the LLM system collected in the post-study questionnaire.

As shown in \autoref{fig:result-1}-d, we observed overall similar task accuracy across conditions without significant difference, \baseline{} (0.657), \tokenUQ{} (0.648), \relationUQ{} (0.646),and \outputUQ{} (0.617).
Contrast to the common expectation, \textbf{providing UQ information does not significantly increase user task accuracy}.
We speculate that this result may stem from two factors. First, uncertainty estimates produced by LLMs are often overconfident~\cite{shorinwa2025survey, yona2024can}, which may limit their effectiveness in calibrating user trust and improve the accuracy of decision-making. 
Second, the medical questions in our study were intentionally designed to be challenging, requiring participants to actively reason with the AI system rather than rely on prior knowledge alone. While this design encouraged engagement, it also posed difficulties for some participants. For example, one participant reported \textit{``carefully verifying information through web search and very confident in their final answers''} in the post-study survey, yet their overall accuracy was only 50\%.








\section{Exploratory Analysis Results}
\label{sec:explore-results}

In addition to confirmatory analyses, we conducted two exploratory investigations: (1) comparing participants' self-confidence with their confidence in AI, and (2) analyzing qualitative feedback from post-study surveys.

\subsection{Confidence in Self versus in LLM}
Although not hypothesized a prior, we observed clear differences between confidence in self (\baseline{}: M=4.253, \outputUQ{}: M=4.094, \relationUQ{}:M=3.962, \tokenUQ{}: M=4.161) and confidence in the LLM (\baseline{}: M=3.578, \outputUQ{}: M=3.498, \relationUQ{}:M=3.442, \tokenUQ{}: M=3.615) . We therefore conducted an exploratory analysis to further examine this pattern. We conducted a mixed-effects analysis with Target (Self vs.\ AI; within-subjects) and Condition (\baseline{}, \outputUQ{}, \tokenUQ{}, \relationUQ{}; between-subjects) as fixed effects, with random intercepts for participants and items.

There was a significant main effect of Target, $\chi^2(1) = 192.43$, $p < .001$, with \textbf{participants reporting higher confidence in their own answers ($M = 4.21$) than in the AI's answers ($M = 3.41$}), a difference of 0.80 points on the 5-point scale. Critically, this effect was moderated by a significant Target $\times$ Condition interaction, $\chi^2(3) = 19.99$, $p < .001$, indicating that \textbf{the self--AI confidence gap varied across uncertainty granularity}.

The self--AI gap was significant in all conditions (all Holm-corrected $p$s $< .001$), but the magnitude differed substantially. The gap was largest in \baseline{} ($\Delta = 1.04$, $SE = 0.08$, $z = 13.87$, $p < .001$), followed by \tokenUQ{} ($\Delta = 0.83$, $SE = 0.08$, $z = 10.22$, $p < .001$) and \outputUQ{} ($\Delta = 0.78$, $SE = 0.09$, $z = 9.01$, $p < .001$), and was smallest in \relationUQ{} ($\Delta = 0.57$, $SE = 0.08$, $z = 7.44$, $p < .001$).
In \baseline{}, confidence in AI was lowest ($M = 3.24$, $SE = 0.07$) while confidence in self was highest ($M = 4.28$, $SE = 0.07$). In contrast, \relationUQ{} showed the highest confidence in AI ($M = 3.52$, $SE = 0.07$) and the lowest confidence in self ($M = 4.09$, $SE = 0.07$). 
The narrowing self-AI confidence gap in the relation-level appears to operate through two mechanisms: increasing perceived AI confidence (through transparent reasoning) and tempering users' self-confidence (by highlighting task complexity and uncertainty). 

\subsection{Qualitative User Feedback}
To further understand the underlying reasons why uncertainty shapes user 
perceptions and behaviors, we conducted a thematic analysis of participants' 
free-form responses to two post-study questions: (1) Describe your experience 
using this interface, and (2) What motivated you to (or not to) use the 
verification tools (links or search)? 
\added{Following Braun and Clarke's analysis framework~\cite{braunclarke2006}, three authors independently familiarized themselves with the data and generated initial codes. Candidate themes were then developed, reviewed against the full dataset, and refined through discussion until consensus was reached. Final themes were assigned descriptive names and synthesized into a coherent account of participants' experiences.}

\subsubsection{Perception of LLM accuracy and uncertainty}

\textbf{Across all conditions, including the \baseline{} condition with no uncertainty information, many participants frequently verified LLM responses because they knew that AI output are not fully accurate.} 
Even though some participants still exhibited over-reliance on AI (\textit{`` I trusted the AI's explanations and did not feel the need to do further verifications''}), many described the AI response as a \textit{starting point} or \textit{staging ground} for further verification.
\textit{``I independently verified each AI answer with my own research''} (P15, \baseline{}).
This behavior may be influenced by the critical nature of medical questions, the widespread awareness of LLM hallucinations, as well as the prominent placement of verification tools in the interface that may increase attention to verification.
For example, participants said they did not believe in AI accuracy because \textit{`` I have used (LLM) before and gotten incorrect responses''}, \textit{`` I have received a lot of inaccurate information (from AI outside of this study)... hesitant for medical research''}.
Additional, broken or irrelevant links in the LLMs responses also made participants realize the LLM hallucination.
\textit{``Some of the links didn't work or weren't available... the AI made a mistake''} (\relationUQ{}).

When uncertainty information was provided, participants exhibited divergent responses across conditions. 
Overall, \outputUQ{} and \relationUQ{} received largely positive feedback. 
In contrast, responses to \tokenUQ{} were more polarized. Some participants found word-level uncertainty cues to function as an \textit{``attention guide''}, helping them identify where additional verification was needed. Others, however, reported that these cues were distracting and reduced readability, characterizing them as \textit{``visual noise''}.

\subsubsection{From Uncertainty Information to Verification Actions}
Participants frequently treated uncertainty as a cue to slow down and cross-check, combined with other cues such as the complexity of the question, inconsistent of the reasoning logic, or the irrelevance and invalidity of the provided links.
For example, one participant in the \outputUQ{} condition noted, \textit{``At around 50\%, I had to really dig to try and fact check the response.''}
A participant from \tokenUQ{} reported ``\textit{When there was a lot of red-marked words… I wanted to double check.}''

At the same time, \textbf{uncertainty is often used as a cue rather than a calibrated indicator of AI reliability}, potentially related to the limited transparency in how uncertainty was generated and concerns about its calibration quality.
First, several participants complained about the lack of transparency in how uncertainty values were generated (\eg, \textit{``I did not understand how the certainty was determined''}), making it hard to make decisions based on uncertainty. 
Second, some participants observed that \textit{``the AI's level of certainty did not always match the quality of its answers''}, reducing their willingness to fully rely their decision on the uncertainty information.

Participants who did not verify responses despite receiving uncertainty information (12 in \relationUQ{}, 5 in \outputUQ{}, and 4 in \tokenUQ{}) reported two primary reasons in the post-study survey. First, some exhibited blind trust in AI (\textit{``I trusted the AI''}, \textit{``AI was enough''}). Second, others expressed high satisfaction with the AI responses, citing their perceived reliability and comprehensive presentation (\textit{``the provided information appeared reliable''}, \textit{``the AI tool summarized all key information''}).

\section{Discussion}

\added{In this section, we discuss the design implications of our findings, followed by the limitations of our study.}

\noindent
\textbf{Fine-Grained Uncertainty May Increase Rather Than Reduce AI Reliance.}
Overreliance is notoriously difficult to mitigate. 
Many methods aim to address this issue by providing additional information such as AI explanations, but were found to be ineffective or even can backfire to increase reliance \cite{bansal2021does, buccinca2021trust}.
While previous studies on AI uncertainty communication showed promising results, these studies primarily examined coarse-grained approaches: single confidence scores or simple natural language expressions \cite{kim2024m, zhang2020effect, rezaeian2025explainability}. 
We show that uncertainty communication at finer granularities, which is increasingly common in LLM UQ, can paradoxically increase user agreement with AI (token-level) or discourage external verification (relation-level) compared with baseline. 
It is important to note that relation-level uncertainty reduced external verification without increase agreement with AI, suggesting that participants may have used the AI uncertainty information as a form of ``internal verification'' and substituting it for external checking.
\added{Our study shows that uncertainty cannot be treated as a simple more-detail-is-better dimension. Instead, research on trustworthy LLM systems should carefully consider how much uncertainty information is presented to users and at what level of granularity.}

\noindent
\added{
\textbf{Uncertainty Granularity Should Be Matched to the Target Behavior.}
Our findings show that uncertainty granularity does not simply produce more or less caution, but selectively shapes which behaviors users engage in. \tokenUQ{} increased agreement without promoting external verification; \relationUQ{} suppressed external verification without increasing agreement; \outputUQ{} reduced users' confidence in their own answers without meaningfully changing either agreement or verification behavior. Each granularity therefore serves a different trust regulation function: token-level cues provide local reassurance that raises overall acceptance; relation-level cues make reasoning transparent enough that users treat the display itself as a substitute for external checking; output-level cues temper users' self-confidence. 
Designers should deliberately align their choice of granularity with the target behavior change, \eg, recalibrate over-confident users, increase AI acceptance, or encourage independent fact-checking.
}

\noindent
\textbf{Critical Thinking Does Not Ensure Improved Performance in Complex Tasks.}
A widely accepted premise in human-AI interaction is that motivating users to engage in critical thinking and verification will improve task performance. Our results challenge this assumption. Even when participants were motivated to verify LLM responses, as evidenced by their report and interaction logs, many still failed to identify and correct errors. 
Task complexity and user ability fundamentally constrain the benefits of critical thinking and verification. 
\added{These results have a direct design implication: for high-complexity domains, uncertainty communication that merely triggers critical thinking is insufficient. It must be paired with explanatory support that actually enables users to understand the verified content.}


\noindent
\textbf{\added{Baseline Skepticism May Attenuate the Benefits of Uncertainty Displays.}}
Contrary to our expectations, we found no significant differences in agreement with AI or external verification between the baseline and certain uncertainty conditions. 
Post-study survey responses revealed surprisingly high awareness of LLM hallucination and the willingness to fact-check, even without explicit uncertainty displays. This baseline skepticism may reflect growing public awareness of LLM limitations, though it may also be influenced by our participant sample, which required at least an undergraduate degree. These findings suggest that the marginal benefit of uncertainty displays may be smaller than anticipated in certain populations.

\noindent
\textbf{Limitations.}
Our findings should be interpreted in light of several limitations. 
First, our crowdsourced study setting may not fully reflect how people use LLMs in everyday contexts, such as personalized information seeking that is driven by individual goals and interests, or complex technical work (\eg, coding) that is not well characterized by binary yes/no decisions. 
Second, the medical questions were intentionally challenging for a general audience to encourage meaningful engagement with AI. But this may have constrained participants' ability to improve performance even when engaging in external verification. 
Consequently, the relationship between uncertainty communication and task accuracy may differ for tasks that better align with users' own expertise or in settings where ground truth is easier to establish. 
Third, we chose did not explicitly calibrate the LLM-generated uncertainty estimates, which can be overconfident and may not reliably track true model reliability~\cite{shorinwa2025survey, yona2024can}. 
We made this choice to realistically reflect the uncertainty signals that current LLM systems might expose, but better-calibrated uncertainty signals could therefore produce different behavioral patterns. 
Finally, our study was conducted in English with U.S.-based participants and restricted to individuals with relatively high educational attainment, which may limit generalizability to other linguistic, cultural, and demographic contexts.

\section{Conclusion}
This study presents a large-scale human-subject experiment examining how uncertainty granularity shapes user reliance, verification behavior, and confidence in LLM-assisted decision making.
%
Our results show that uncertainty granularity is a consequential design decision that shapes reliance, verification behavior, and user confidence in LLM-assisted decision making. 
Token-level uncertainty increased overall agreement with the AI. In contrast, both relation-level and output-level uncertainty reduced users’ confidence in their own final answers, whereas token-level uncertainty did not.
In addition, relation-level uncertainty reduced external verification (link clicking and web search) and increased self-reported reliance on the AI (including its uncertainty information), without increasing agreement.
Taken together, these results suggest that fine-grained cues (token-level) can increase acceptance of AI outputs without improving users’ ability to detect errors, while coarser uncertainty presentations may temper users’ self-confidence. Moreover, the reduced verification observed in the relation-level condition indicates a potential design risk: structured uncertainty representations may encourage users to rely on the AI provided uncertainty information rather than seeking independent external verification.



\section*{Generative AI Usage Statement}
Generative AI (ChatGPT 5.2, Claude Sonnet 4.5) were used to assist with grammar editing, style refinement, and to provide feedback on various titles proposed by the authors. No original ideas or data were generated by AI. The authors take full responsibility for the accuracy and integrity of this work.

\bibliographystyle{ACM-Reference-Format}
\bibliography{reference}

\appendix
\section*{APPENDIX}
The appendix is structured in the following way.
\begin{itemize}
    \item \autoref{sup:participants}: Participant Demographics and Background
    \item \autoref{sup:questions}: Questions Used in the In-Study Tasks and the Post-Study Questionnaires
    \item \autoref{sup:interface}:  Task Interface and LLM Responses
    
\end{itemize}

\section{Participant Demographics and Background}
\label{sup:participants}

\added{
Participants were recruited through Prolific, a widely used crowdsourcing platform for behavioral research, and screened for U.S. residency in accordance with our approved IRB protocol. To ensure data quality, we applied the following eligibility criteria: a Prolific approval rating of at least 95\%, basic familiarity with data analysis, an undergraduate degree or higher, prior experience with LLMs, and no medical occupation or domain expertise. Participants were then randomly assigned to one of four conditions: a no-uncertainty baseline and three uncertainty granularity conditions. Demographic information was collected via an intake form (see \autoref{tab:demographics}), reflecting a diverse sample. No significant differences in demographic distributions or prior AI experience were found across conditions (one-way ANOVA; AI usage onset: $p = .693$; AI usage frequency: $p = .937$).}

\renewcommand{\arraystretch}{0.65} 
\begin{table}[H]
  \centering
  \caption{Participant demographics (final sample; $N=192$).}
  \vspace{-1em}
  \label{tab:demographics}
  \small
  \setlength{\tabcolsep}{6pt}
  \begin{tabular}{@{} p{0.28\linewidth} p{0.46\linewidth} r r @{}}
    \toprule
    \textbf{Variable} & \textbf{Category} & \textbf{n} & \textbf{\%} \\
    \midrule
    Age & 18--24 & 7 & 3.8 \\
        & 25--34 & 57 & 31.1 \\
        & 35--44 & 52 & 28.4 \\
        & 45--54 & 35 & 19.1 \\
        & 55--64 & 25 & 13.7 \\
        & 65+ & 6 & 3.3 \\
    \addlinespace

    Education & Bachelor's degree & 125 & 68.3 \\
              & Master's degree & 34 & 18.6 \\
              & Doctoral degree & 16 & 8.7 \\
              & Associate's degree & 4 & 2.2 \\
              & Some college & 2 & 1.1 \\
    \addlinespace

    Time since first LLM use & 1--2 years ago & 76 & 41.5 \\
                             & More than 2 years ago & 62 & 33.9 \\
                             & 6--12 months ago & 32 & 17.5 \\
                             & Within the last 6 months & 11 & 6.0 \\
    \addlinespace

    LLM usage frequency & Daily & 98 & 53.6 \\
                        & A few times a week & 53 & 29.0 \\
                        & A few times a month & 24 & 13.1 \\
                        & Never & 4 & 2.2 \\
    \addlinespace

    LLM use cases$^\dagger$ & Searching for information & 159 & 86.9 \\
                            & Writing or editing text & 124 & 67.8 \\
                            & Studying or learning & 112 & 61.2 \\
                            & Data analysis / research & 111 & 60.7 \\
                            & Creative tasks & 106 & 57.9 \\
                            & Coding / technical work & 53 & 29.0 \\
    \bottomrule
  \end{tabular}

  \vspace{2pt}
  \footnotesize{$^\dagger$ Multiple selections allowed; percentages may sum to more than 100\%.}
\end{table}

\newpage
\section{Questions Used in the In-Study Tasks and the Post-Study Questionnaires}
\label{sup:questions}

\subsection{Questions Used in the In-Study Tasks}
\begin{enumerate}[leftmargin=*, label=\textbf{\arabic*.}, itemsep=1.2em]

\item What is your final answer?

\noindent
\radio\ Yes \hspace{2.0em} \radio\ No

\item How confident are you in the AI's answer?

\noindent
\RadioScaleFive

\item How confident are you in your own answer?

\noindent
\RadioScaleFive

\item What was your final answer based on? (Select all that apply)

\begin{itemize}[leftmargin=1.6em, itemsep=0.4em, topsep=0.2em]
  \item \checkbox\ AI system's answer
  \item \checkbox\ Linked sources in the AI answer
  \item \checkbox\ Your own Internet search
\end{itemize}

\end{enumerate}


\subsection{Post-Study Survey}

    


\noindent\textit{Please indicate how much you agree with the following statements. (1 = Strongly disagree, 5 = Strongly agree).}

\vspace{0.5em}

\begin{enumerate}
[
  leftmargin=*,
  label=\textbf{\arabic*.},
  itemsep=0.8em
]

  \item The interface seemed competent at answering the medical questions.
  \LikertFive

  \item I would be willing to use this interface again for similar medical questions.
  \LikertFive

  \item The AI system felt human-like.
  \LikertFive

  \item I felt I had a good understanding of what the AI system's answers were based on.
  \LikertFive

  \item I felt I had a good understanding of when the AI system's answers might be wrong.
  \LikertFive

  \item Please describe your experience using this interface.\\
  \textit{You may comment on anything you found helpful, confusing, surprising, frustrating, or interesting.}
  \FreeResponseBox[0.5cm]

  \item If you used the validation tools (links or search), what motivated you to do so? If you did not, why not?
  \FreeResponseBox[0.5cm]

\end{enumerate}



\section{Task Interface and LLM Responses}
\label{sup:interface}


\subsection{Correct Answers across LLM Responses}
\label{sup:uq_distribution}
 
Table~\ref{tab:ai_correctness} shows the AI answer and \added{the correct answer} for each of the 
eight questions used in the study.
 
\begin{table}[H]
\centering
\caption{AI correctness \added{and correct answer} for each question.}
\label{tab:ai_correctness}
\small
\begin{tabular}{@{} p{0.55\linewidth} c c @{}}
\toprule
\textbf{Question} & \textbf{AI Answer} & \added{\textbf{Correct Answer}} \\
\midrule
Did dupilumab receive FDA approval for asthma before chronic rhinosinusitis? & No & \added{Yes} \\
Is there more antihistamine in Benadryl than Rhinocort? & No & \added{Yes} \\
Is deep vein thrombosis a common side effect of Ocella? & Yes & \added{No} \\
Is spironolactone an FDA-approved drug for treating acne? & Yes & \added{No} \\
Are both Simvastatin and Ambien recommended to be taken at night? & Yes & \added{Yes} \\
Is uveitis a common symptom of ankylosing spondylitis? & Yes & \added{Yes} \\
Is fever a common symptom of jock itch? & No & \added{No} \\
Can an adult who has not had chickenpox get shingles? & No & \added{No} \\
\bottomrule
\end{tabular}
\end{table}
 

\subsection{LLM Responses}
\subsection*{\added{Q1: Did Dupilumab receive FDA approval for asthma before chronic rhinosinusitis?}}
\addcontentsline{toc}{subsection}{Q1: Dupilumab FDA approval}

\added{The sequence of FDA approvals for Dupilumab indicates that its approval for asthma occurred after its approval for chronic rhinosinusitis. This suggests that the regulatory pathway for chronic rhinosinusitis was prioritized, possibly due to the pressing need for effective treatments in that area, which was supported by clinical evidence demonstrating significant benefits for patients suffering from chronic rhinosinusitis [1]. Furthermore, the approval timeline reflects a strategic approach by the manufacturers to address the most urgent health concerns first, which can often influence the order of approvals [2].}

\added{Additionally, the approval for chronic rhinosinusitis was based on a robust set of clinical trials that highlighted the drug's efficacy in managing symptoms and improving quality of life for patients. This foundational approval likely set the stage for subsequent applications for other indications, such as asthma, which may have required additional data to support its use in that context [1]. The regulatory decisions made in this case illustrate the complexities involved in drug approvals, where the urgency of treatment needs can dictate the order in which approvals are granted [2].}

\subsection*{\added{Q2: Is there more antihistamine in Benadryl than Rhinocort?}}
\addcontentsline{toc}{subsection}{Q2: Benadryl vs.\ Rhinocort antihistamine}

\added{The analysis suggests that Benadryl does not contain more antihistamine than Rhinocort based on their distinct pharmacological compositions. Rhinocort, primarily formulated as a corticosteroid nasal spray, exerts downstream antihistamine-like effects through anti-inflammatory mechanisms that are more concentrated and targeted for allergic reactions than the antihistamine present in Benadryl. This specialized formulation in Rhinocort allows it to deliver a higher effective dose of antihistamine activity despite Benadryl's reputation as an antihistamine medication [1].}

\added{However, one could argue that the difference in delivery methods and drug classes complicates direct comparisons of antihistamine content. Benadryl, an oral antihistamine, contains a different class of antihistamine agents that act systemically but at lower concentrations compared to Rhinocort's localized and potent antihistamine effect. This raises a concern that simply measuring total antihistamine content without considering pharmacodynamics and administration routes may misrepresent the actual antihistamine potency, supporting the conclusion that Benadryl does not have more antihistamine than Rhinocort [2].}

\subsection*{\added{Q3: Is deep vein thrombosis a common side effect of Ocella?}}
\addcontentsline{toc}{subsection}{Q3: DVT and Ocella}

\added{Deep Vein Thrombosis (DVT) is a common side effect of Ocella due to its hormonal composition, which can increase blood coagulability and promote clot formation in susceptible individuals. The estrogen component in Ocella is known to elevate the risk of thrombotic events by altering the balance of clotting factors, making DVT a frequently observed adverse effect in patients using this medication [1]. This risk is particularly notable in women with additional predisposing factors such as obesity or prolonged immobility, which further compounds the likelihood of developing DVT while on Ocella therapy.}

\added{Clinical observations and pharmacological analyses have consistently demonstrated that Ocella's formulation contributes to a prothrombotic state, thereby increasing the incidence of DVT among its users compared to non-users [2]. The mechanism involves enhanced platelet aggregation and reduced fibrinolysis, which together facilitate venous clot formation. These findings underscore the importance of monitoring patients on Ocella for signs of DVT and considering risk factors before prescribing this medication [2].}

\subsection*{\added{Q4: Is spironolactone an FDA-approved drug for treating acne?}}
\addcontentsline{toc}{subsection}{Q4: Spironolactone for acne}

\added{Spironolactone is an FDA-approved drug for treating acne, as it effectively reduces androgen levels that contribute to acne development. Its anti-androgenic properties help decrease sebum production, which is a primary factor in acne formation, making it a recognized treatment option for hormonal acne in particular [1]. This approval reflects its established role in dermatology for managing acne, especially in adult women where hormonal influences are significant.}

\added{Clinical evidence supports spironolactone's efficacy and safety profile in acne treatment, which has led to its formal approval by regulatory authorities. The drug's ability to target the hormonal pathways involved in acne pathogenesis distinguishes it from other treatments, providing a valuable alternative for patients who do not respond well to conventional therapies [2]. This regulatory endorsement confirms its therapeutic value and widespread acceptance in acne management protocols.}

\subsection*{\added{Q5: Are both Simvastatin and Ambien drugs recommended to be taken at night?}}
\addcontentsline{toc}{subsection}{Q5: Simvastatin and Ambien timing}

\added{Both Simvastatin and Ambien are recommended to be taken at night due to their pharmacological profiles and intended effects. Simvastatin, a lipid-lowering agent, is most effective when taken at night because cholesterol synthesis in the liver peaks during this time, enhancing the drug's ability to reduce LDL cholesterol levels. Ambien, a sedative-hypnotic used for insomnia, is naturally advised for nighttime use to align with the sleep cycle and promote effective sleep onset and maintenance [1]. This timing optimizes the therapeutic benefits of both medications by synchronizing with the body's biological rhythms.}

\added{Taking Simvastatin and Ambien at night also minimizes potential side effects and maximizes patient compliance. Nighttime administration of Simvastatin reduces the risk of muscle-related side effects by allowing the body to metabolize the drug during rest periods, while Ambien's sedative effects are best utilized when the patient is preparing for sleep, reducing daytime drowsiness and cognitive impairment. Clinical guidelines and pharmacokinetic studies support this dual recommendation, confirming that nighttime dosing is the optimal strategy for both drugs [2].}

\subsection*{\added{Q6: Is uveitis a common symptom of ankylosing spondylitis?}}
\addcontentsline{toc}{subsection}{Q6: Uveitis and ankylosing spondylitis}

\added{Uveitis is a common symptom of Ankylosing Spondylitis due to the inflammatory nature of the disease, which frequently affects the eyes alongside the spine and joints. The immune system's dysregulation in Ankylosing Spondylitis often leads to inflammation in the uveal tract, causing symptoms such as eye pain, redness, and blurred vision. This ocular involvement is well-documented and considered a hallmark extra-articular manifestation, supporting the strong association between uveitis and Ankylosing Spondylitis [1]. The recurrent episodes of uveitis in patients with this condition highlight its prevalence and clinical significance in disease management.}

\subsection*{\added{Q7: Is fever a common symptom of jock itch?}}
\addcontentsline{toc}{subsection}{Q7: Fever and jock itch}

\added{Fever is not a common symptom of Jock Itch, as this condition primarily affects the skin and is characterized by localized itching, redness, and irritation rather than systemic symptoms like fever. The fungal infection responsible for Jock Itch typically remains confined to the groin area, causing discomfort and rash without triggering the body's systemic inflammatory response that would result in fever [1]. This localized nature of the infection means that fever is rarely observed in patients suffering from Jock Itch, distinguishing it from other infections that involve deeper tissues or systemic involvement.}

\added{Although some infections can cause fever due to widespread inflammation, Jock Itch's superficial fungal infection does not usually provoke such a response. However, a competing interpretation suggests that if the infection becomes severe or complicated by a secondary bacterial infection, fever might occur, but this is an exception rather than the rule [2]. Therefore, while fever can theoretically arise in rare complicated cases, it is not a typical or common symptom of Jock Itch, reinforcing the conclusion that fever is generally absent in this condition.}

\subsection*{\added{Q8: Can an adult who has not had chickenpox get shingles?}}
\addcontentsline{toc}{subsection}{Q8: Chickenpox and shingles}

\added{An adult who has not had chickenpox cannot develop shingles because shingles arises from the reactivation of the varicella-zoster virus that remains dormant in nerve cells after an initial chickenpox infection. Without a prior chickenpox infection, the virus is not present in the body to reactivate later as shingles, making it impossible for shingles to occur in someone who has never had chickenpox [1]. This biological mechanism clearly indicates that shingles is contingent upon a previous exposure to the chickenpox virus, which establishes the necessary viral latency for shingles to develop.}

\added{However, it is important to note that while chickenpox infection is a prerequisite for shingles, vaccination against chickenpox can also introduce a weakened form of the virus that may later reactivate as shingles. Yet, in adults who have neither had chickenpox nor received the vaccine, the absence of the virus means shingles cannot manifest. This distinction underscores that shingles strictly depends on the presence of the varicella-zoster virus from a prior infection or vaccination, and without such exposure, shingles cannot occur [2].}

\subsection{Uncertainty Interface}
\subsection*{Q1: Did Dupilumab receive FDA approval for asthma before chronic rhinosinusitis?}
 
\begin{figure}[H]
    \centering
    \includegraphics[width=\linewidth]{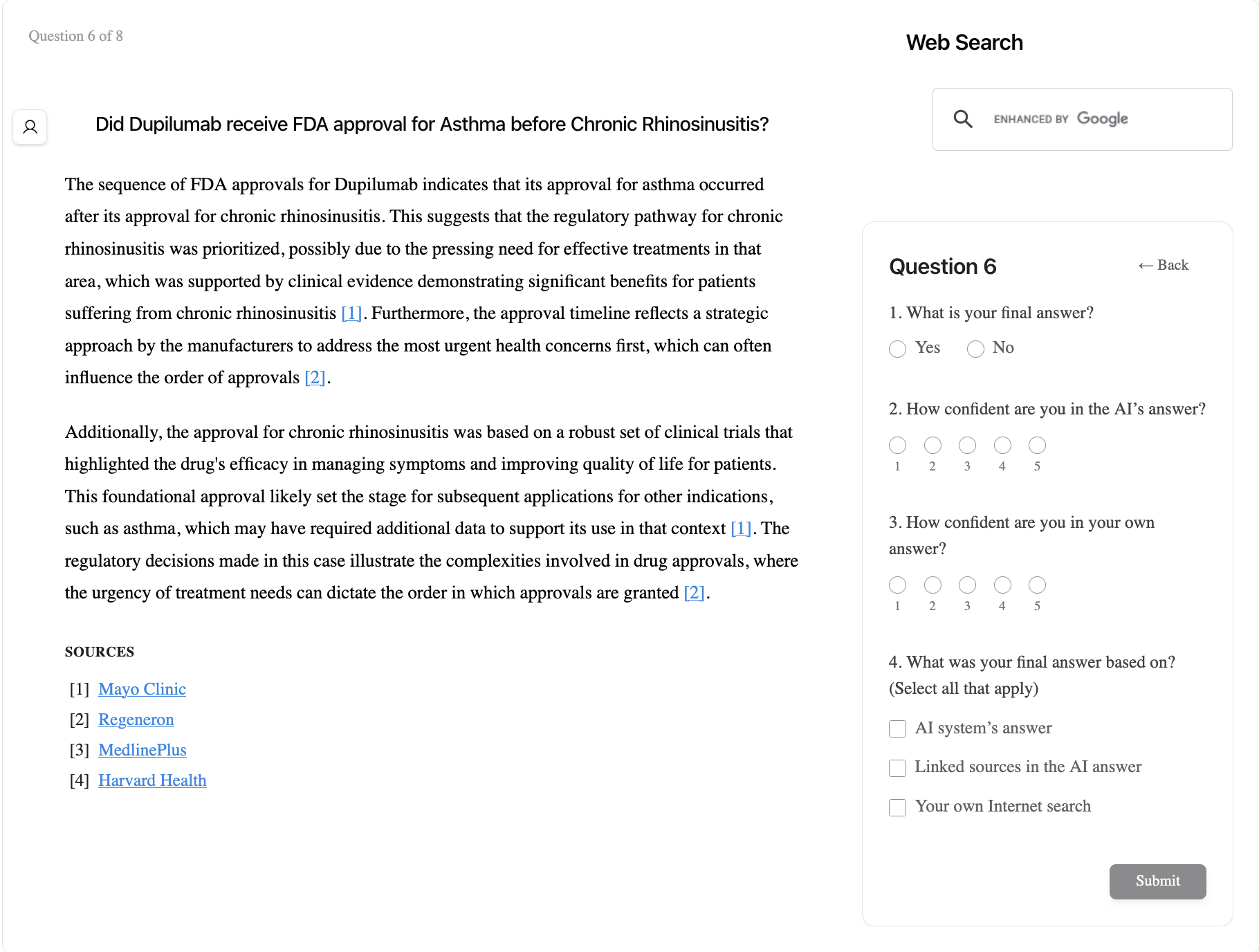}
    \caption{Q1 -- Baseline condition.}
    \Description{Screenshot of the task interface for question 1 (Did Dupilumab receive FDA approval for asthma before chronic rhinosinusitis?) in the Baseline condition, showing the LLM's two-paragraph response with reference links but no uncertainty information displayed.}
    \label{fig:q1_baseline}
\end{figure}
 
\begin{figure}[H]
    \centering
    \includegraphics[width=\linewidth]{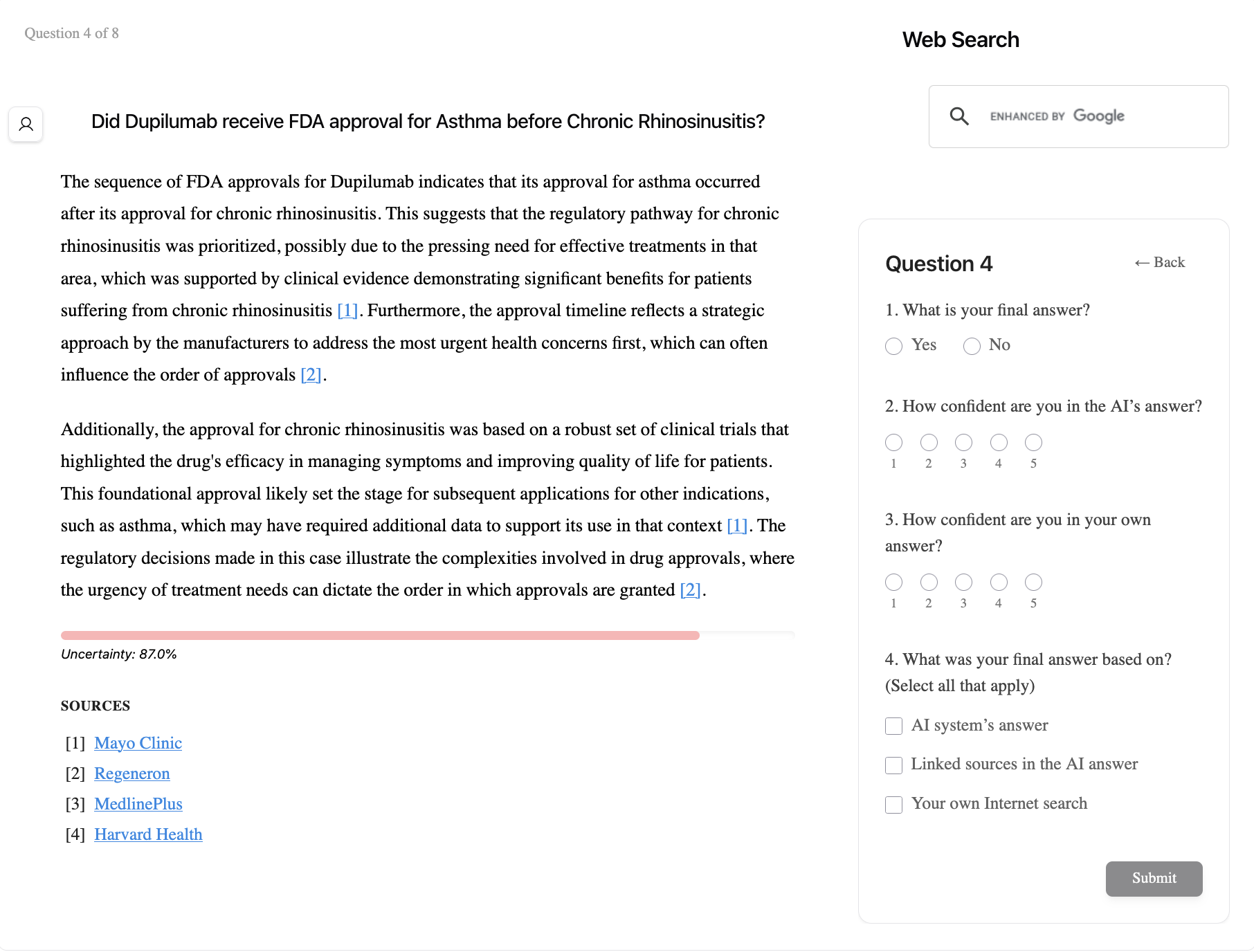}
    \caption{Q1 -- Output-Level UQ condition.}
    \Description{Screenshot of the task interface for question 1 in the Output-Level UQ condition, showing the LLM's two-paragraph response with a single horizontal uncertainty bar and percentage displayed beneath the response.}
    \label{fig:q1_output}
\end{figure}
 
\begin{figure}[H]
    \centering
    \includegraphics[width=\linewidth]{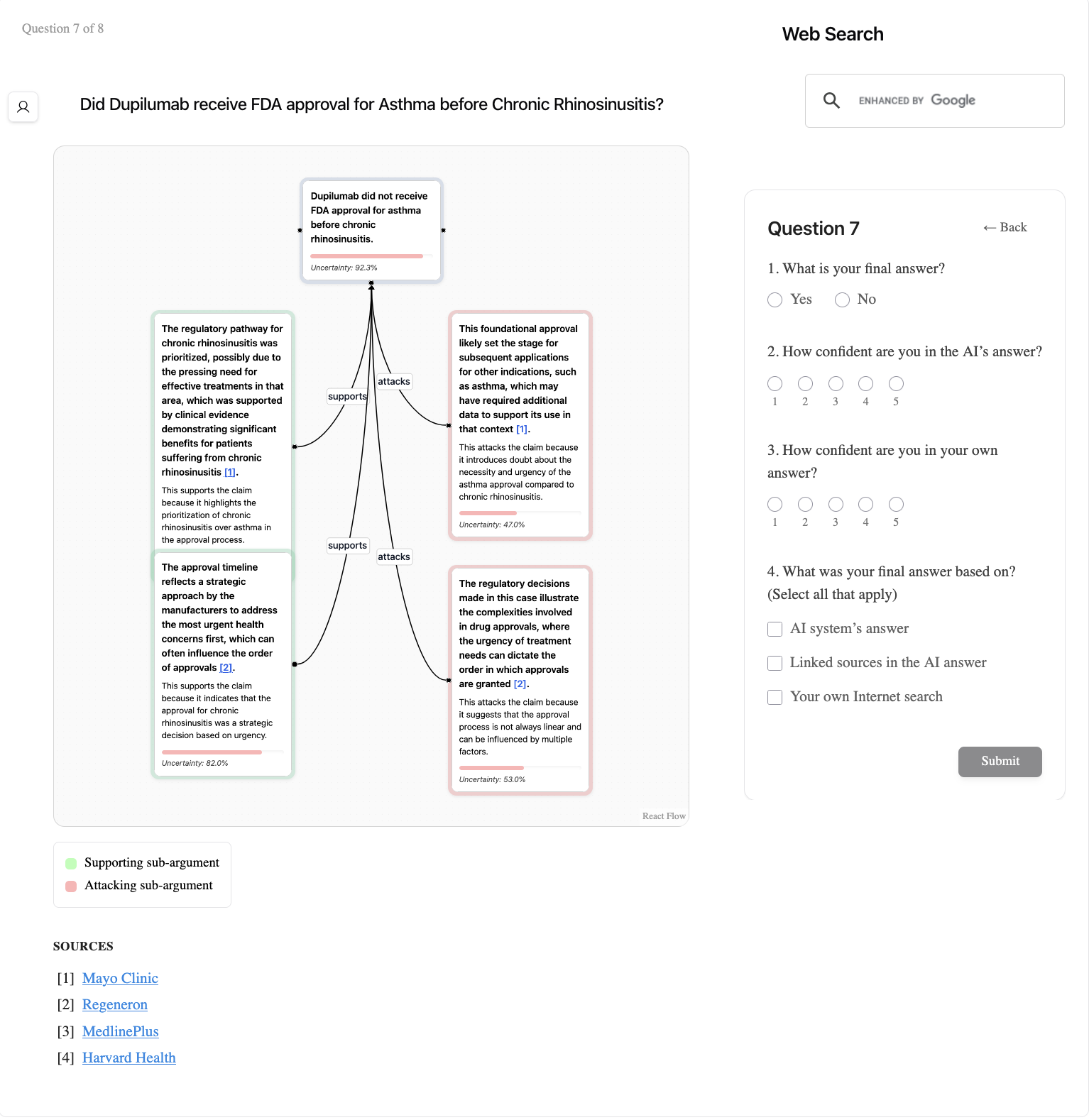}
    \caption{Q1 -- Relation-Level UQ condition.}
    \Description{Screenshot of the task interface for question 1 in the Relation-Level UQ condition, showing the LLM's response decomposed into a node-link diagram with a central claim and four sub-arguments (two supporting, two attacking), each annotated with its own uncertainty score.}
    \label{fig:q1_relation}
\end{figure}
 
\begin{figure}[H]
    \centering
    \includegraphics[width=\linewidth]{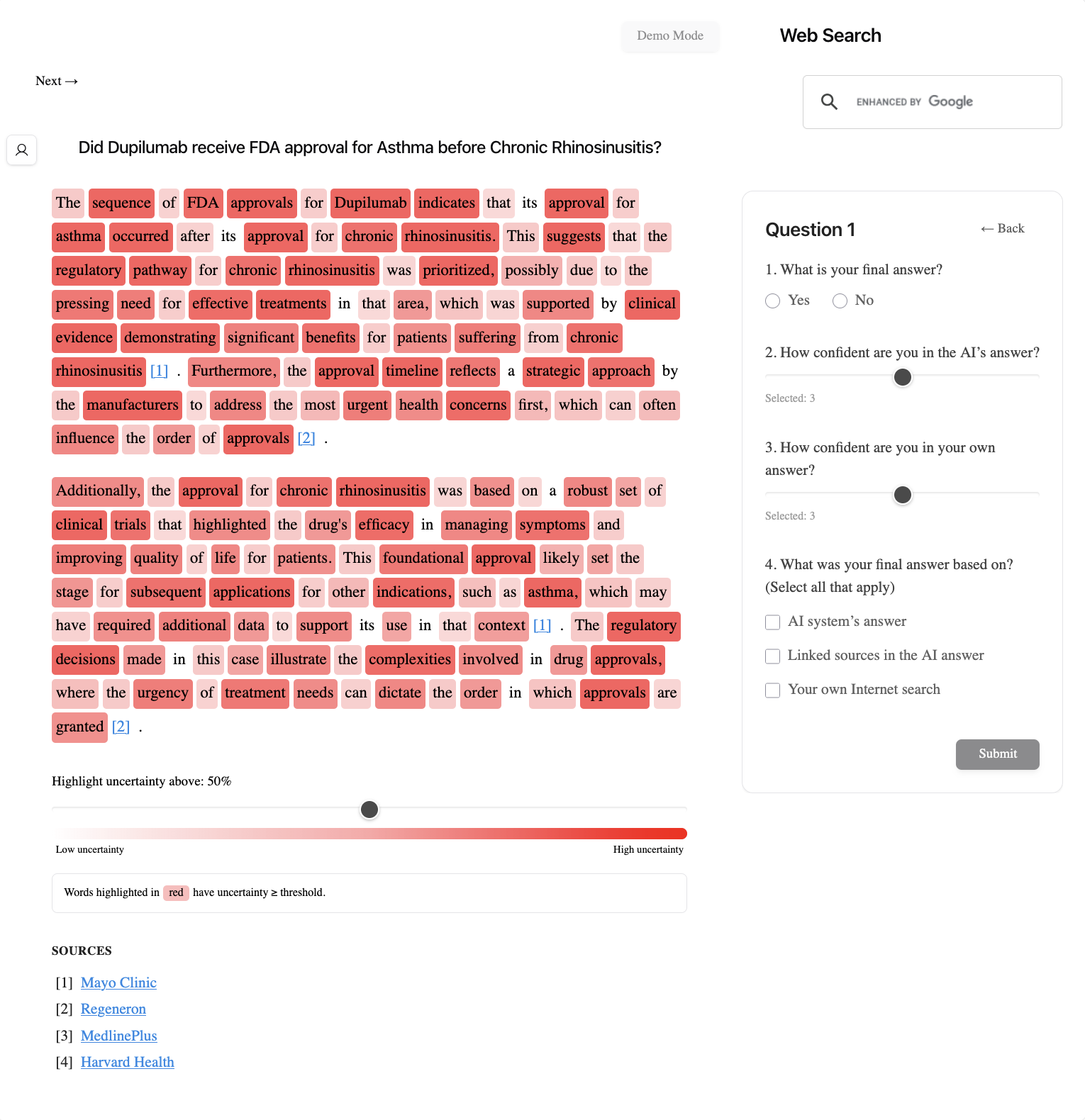}
    \caption{Q1 -- Token-Level UQ condition.}
    \Description{Screenshot of the task interface for question 1 in the Token-Level UQ condition, showing the LLM's two-paragraph response with individual words highlighted in red at varying intensities according to their uncertainty scores, plus an interactive threshold slider.}
    \label{fig:q1_token}
\end{figure}
 
\subsection*{Q2: Is there more antihistamine in Benadryl than Rhinocort?}
 
\begin{figure}[H]
    \centering
    \includegraphics[width=\linewidth]{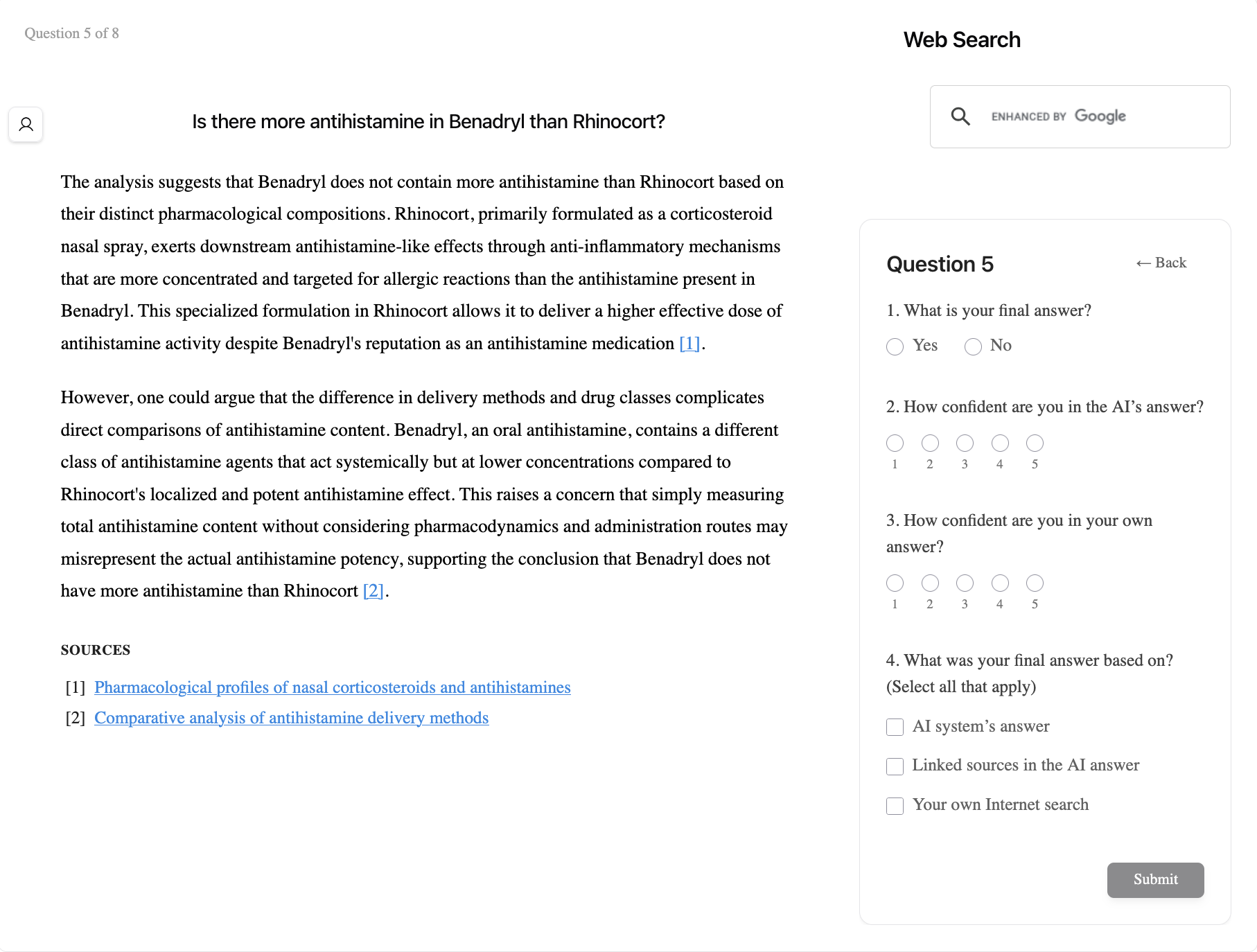}
    \caption{Q2 -- Baseline condition.}
    \Description{Screenshot of the task interface for question 2 (Is there more antihistamine in Benadryl than Rhinocort?) in the Baseline condition, showing the LLM's two-paragraph response with reference links but no uncertainty information displayed.}
    \label{fig:q2_baseline}
\end{figure}
 
\begin{figure}[H]
    \centering
    \includegraphics[width=\linewidth]{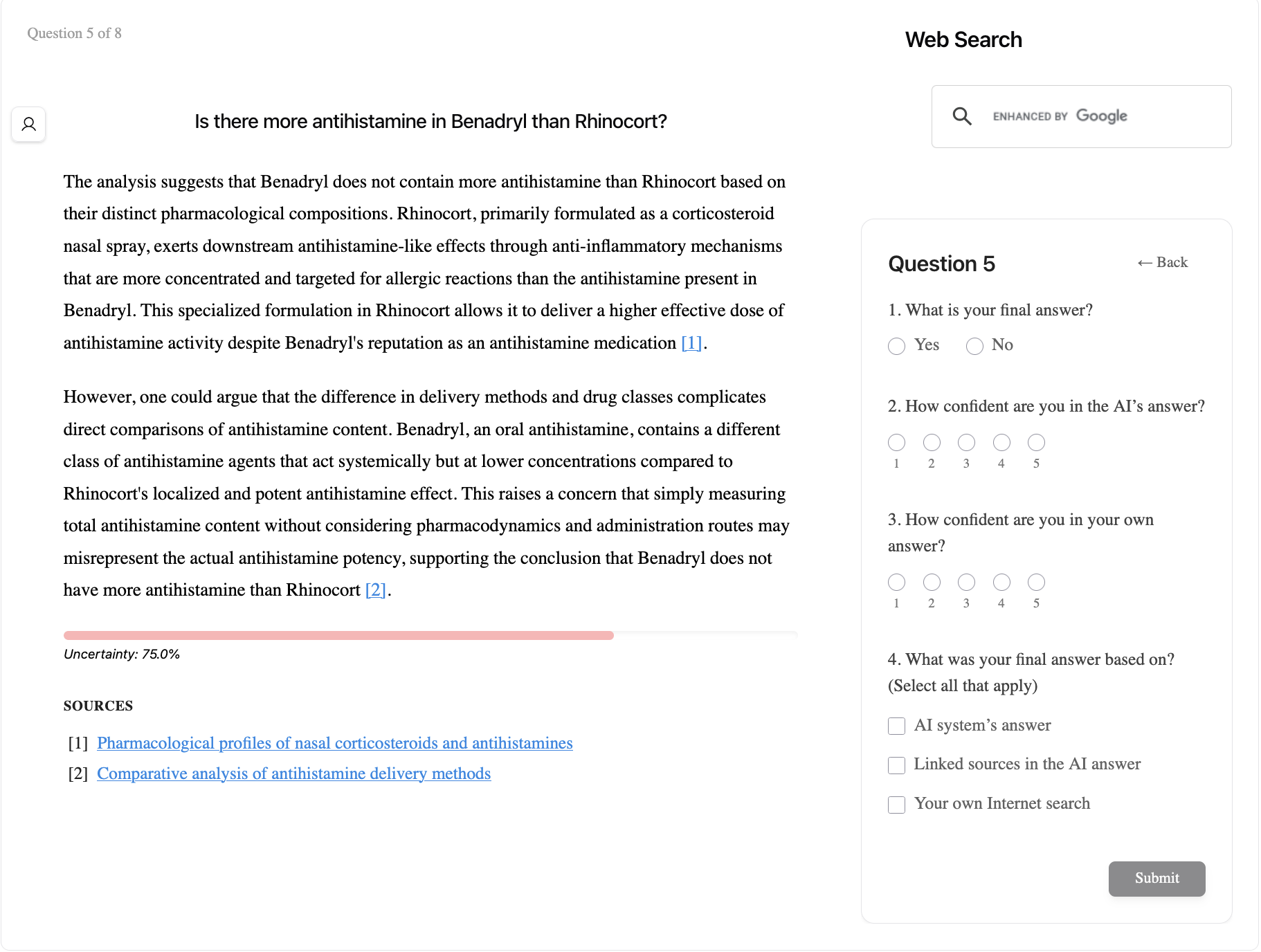}
    \caption{Q2 -- Output-Level UQ condition.}
    \Description{Screenshot of the task interface for question 2 in the Output-Level UQ condition, showing the LLM's two-paragraph response with a single horizontal uncertainty bar and percentage displayed beneath the response.}
    \label{fig:q2_output}
\end{figure}
 
\begin{figure}[H]
    \centering
    \includegraphics[width=\linewidth]{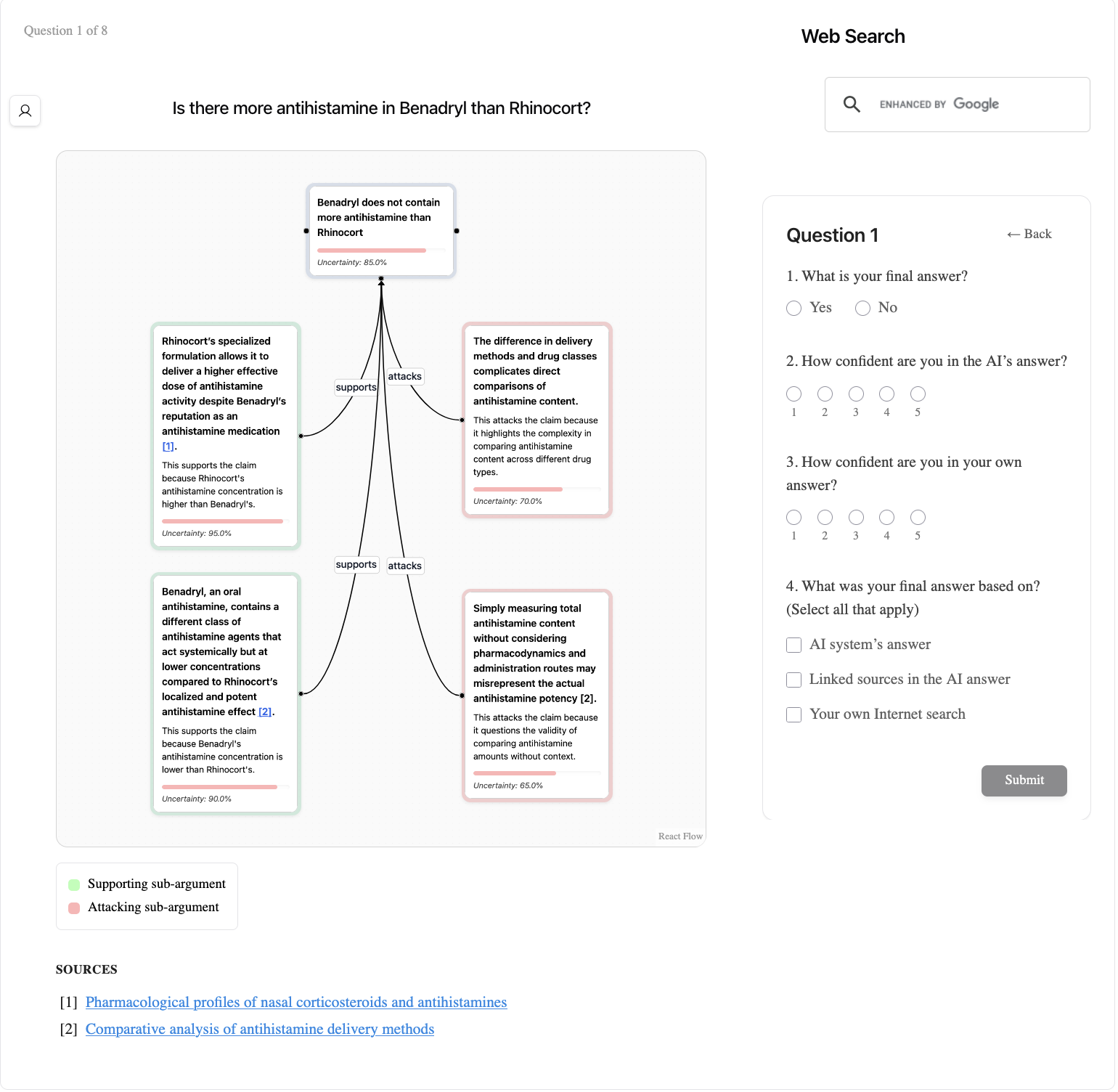}
    \caption{Q2 -- Relation-Level UQ condition.}
    \Description{Screenshot of the task interface for question 2 in the Relation-Level UQ condition, showing the LLM's response decomposed into a node-link diagram with a central claim and four sub-arguments (two supporting, two attacking), each annotated with its own uncertainty score.}
    \label{fig:q2_relation}
\end{figure}
 
\begin{figure}[H]
    \centering
    \includegraphics[width=\linewidth]{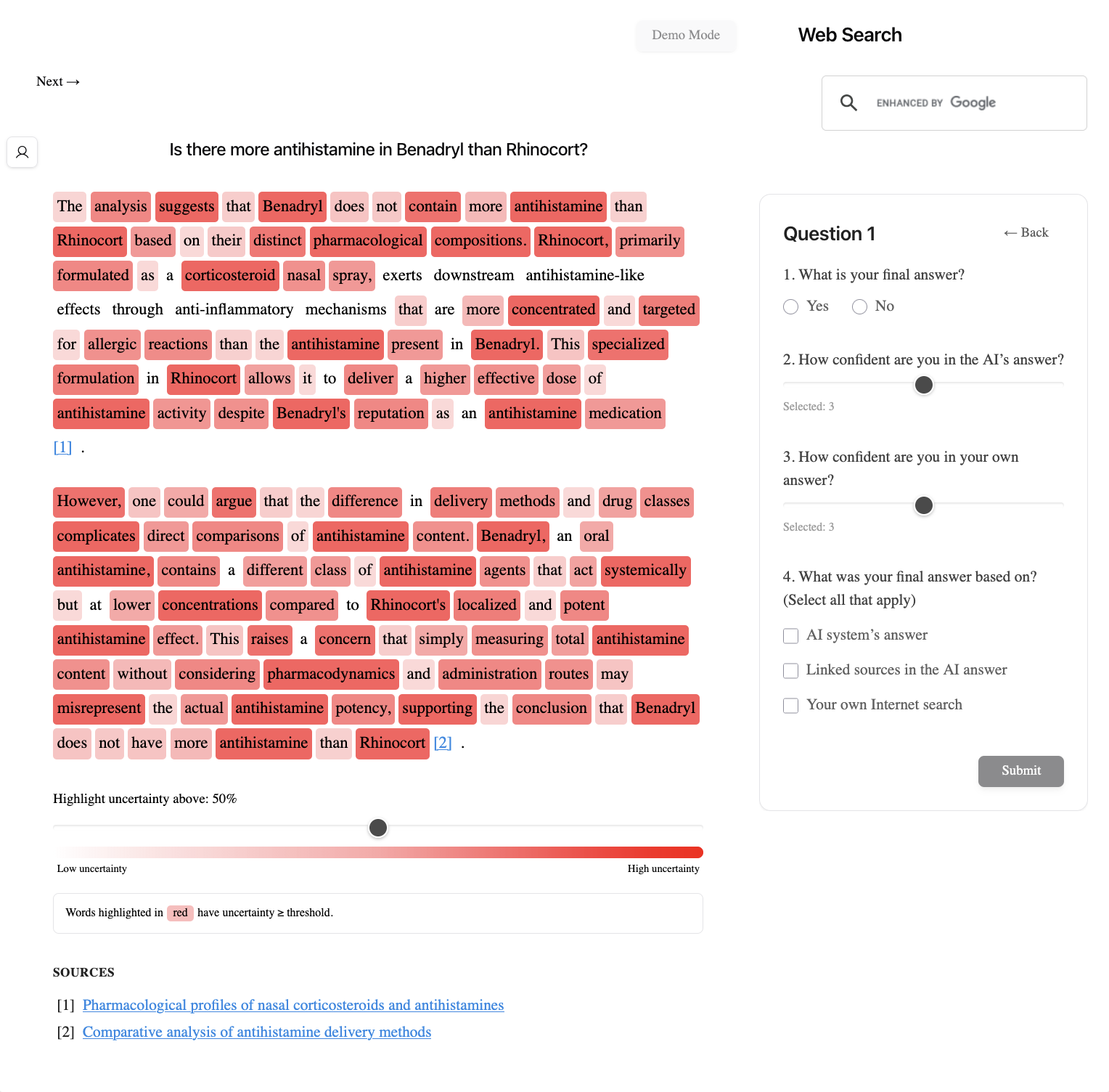}
    \caption{Q2 -- Token-Level UQ condition.}
    \Description{Screenshot of the task interface for question 2 in the Token-Level UQ condition, showing the LLM's two-paragraph response with individual words highlighted in red at varying intensities according to their uncertainty scores, plus an interactive threshold slider.}
    \label{fig:q2_token}
\end{figure}
 
\subsection*{Q3: Is deep vein thrombosis a common side effect of Ocella?}
 
\begin{figure}[H]
    \centering
    \includegraphics[width=\linewidth]{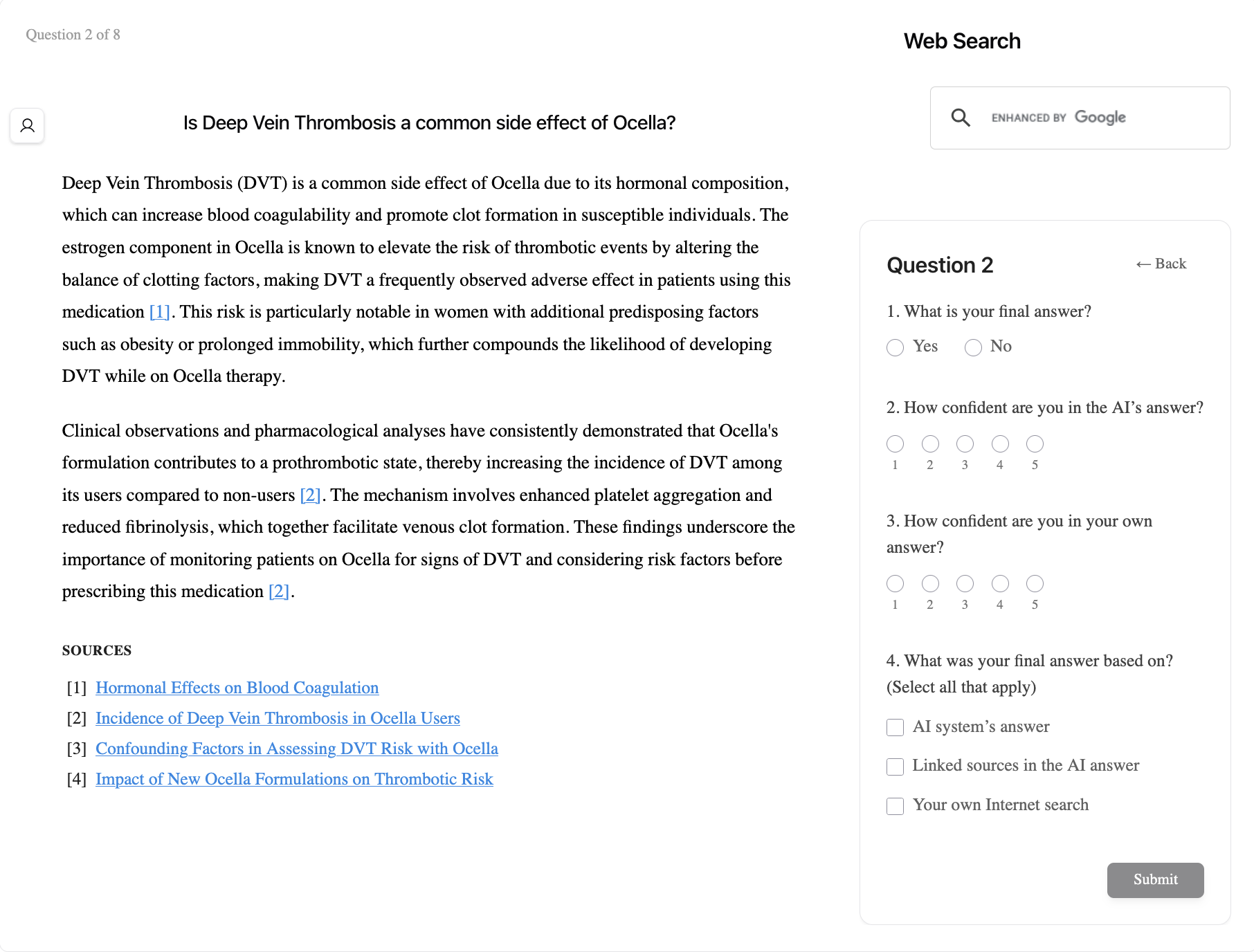}
    \caption{Q3 -- Baseline condition.}
    \Description{Screenshot of the task interface for question 3 (Is deep vein thrombosis a common side effect of Ocella?) in the Baseline condition, showing the LLM's two-paragraph response with reference links but no uncertainty information displayed.}
    \label{fig:q3_baseline}
\end{figure}
 
\begin{figure}[H]
    \centering
    \includegraphics[width=\linewidth]{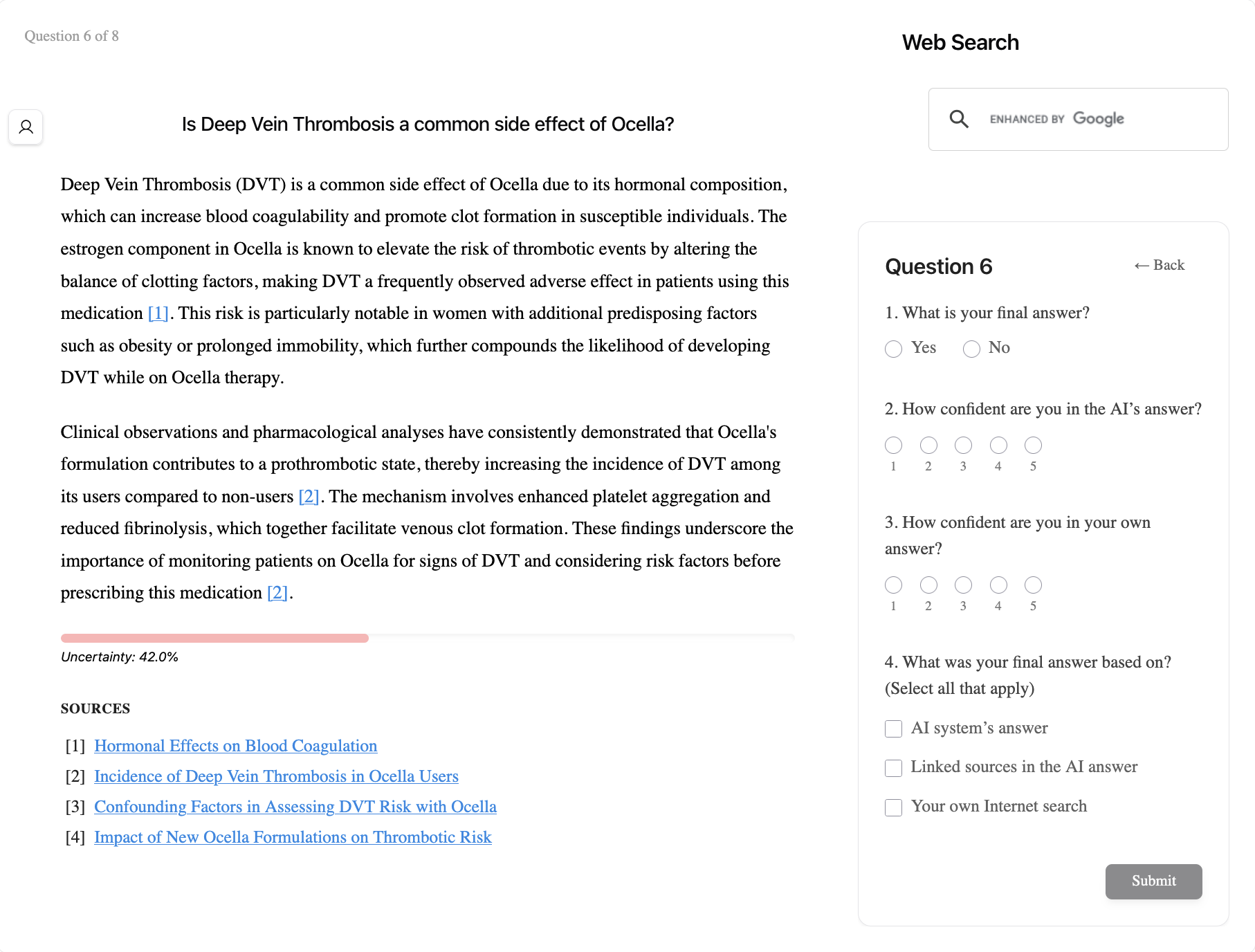}
    \caption{Q3 -- Output-Level UQ condition.}
    \Description{Screenshot of the task interface for question 3 in the Output-Level UQ condition, showing the LLM's two-paragraph response with a single horizontal uncertainty bar and percentage displayed beneath the response.}
    \label{fig:q3_output}
\end{figure}
 
\begin{figure}[H]
    \centering
    \includegraphics[width=\linewidth]{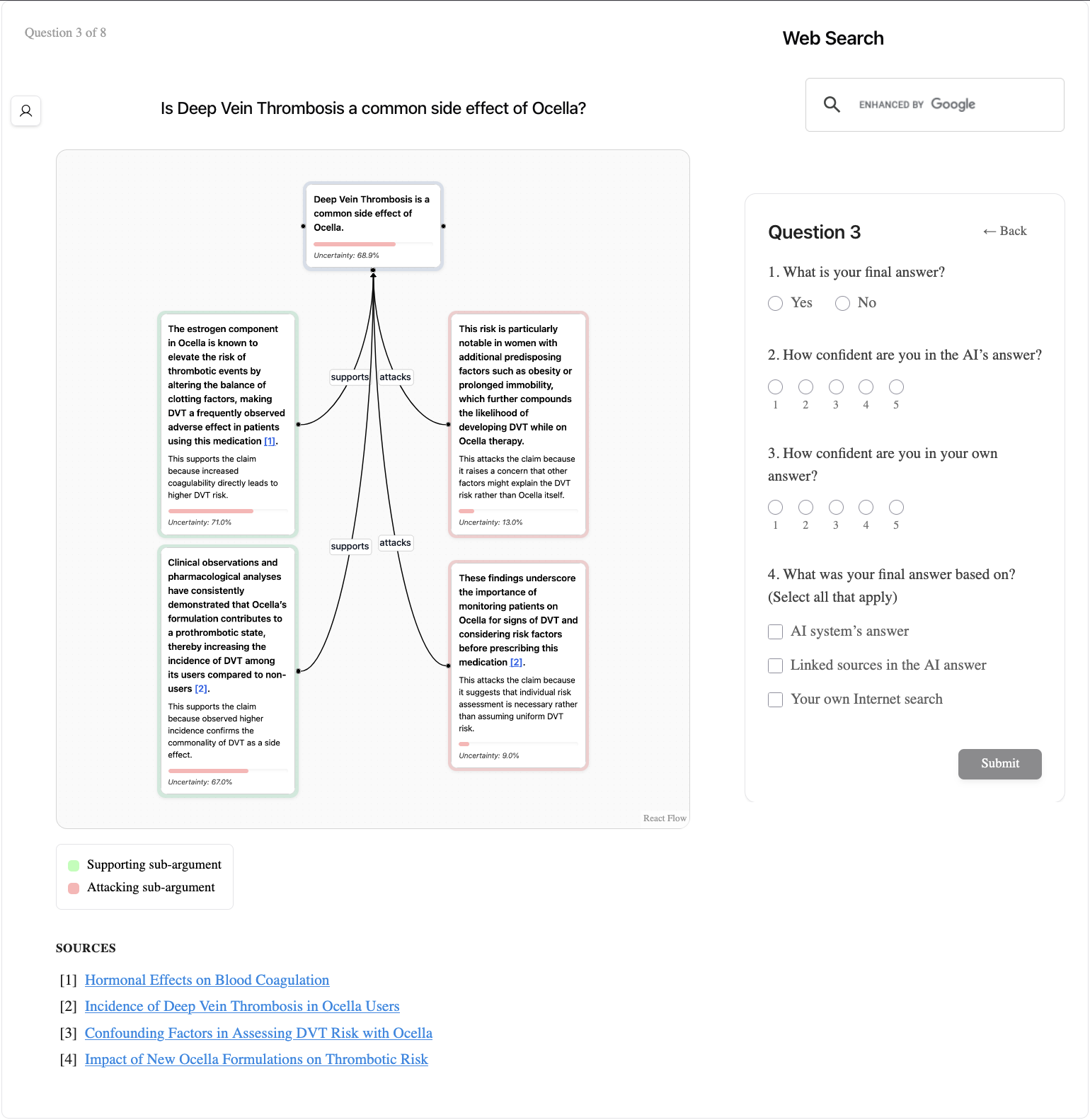}
    \caption{Q3 -- Relation-Level UQ condition.}
    \Description{Screenshot of the task interface for question 3 in the Relation-Level UQ condition, showing the LLM's response decomposed into a node-link diagram with a central claim and four sub-arguments (two supporting, two attacking), each annotated with its own uncertainty score.}
    \label{fig:q3_relation}
\end{figure}
 
\begin{figure}[H]
    \centering
    \includegraphics[width=\linewidth]{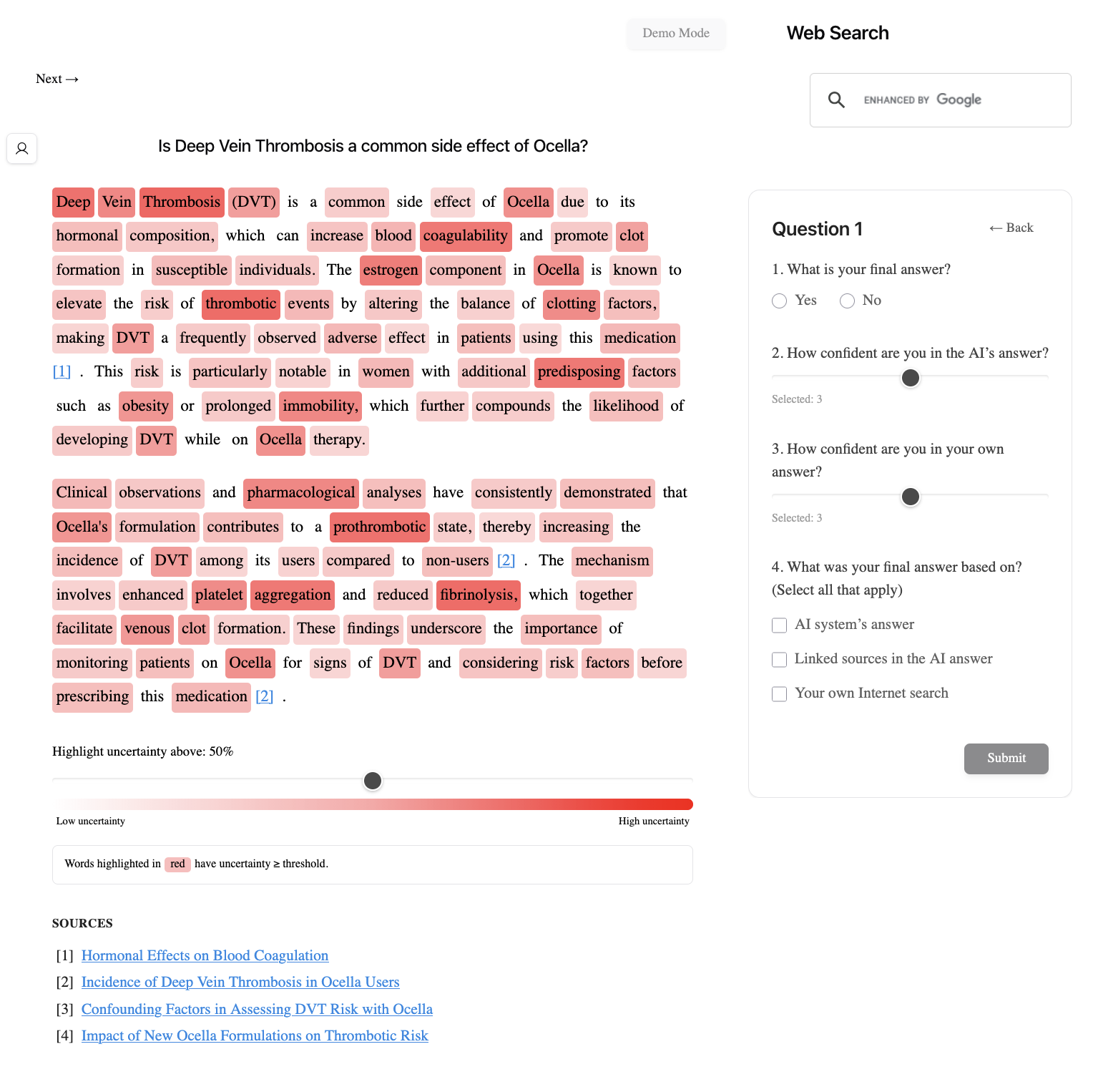}
    \caption{Q3 -- Token-Level UQ condition.}
    \Description{Screenshot of the task interface for question 3 in the Token-Level UQ condition, showing the LLM's two-paragraph response with individual words highlighted in red at varying intensities according to their uncertainty scores, plus an interactive threshold slider.}
    \label{fig:q3_token}
\end{figure}
 
\subsection*{Q4: Is spironolactone an FDA-approved drug for treating acne?}
 
\begin{figure}[H]
    \centering
    \includegraphics[width=\linewidth]{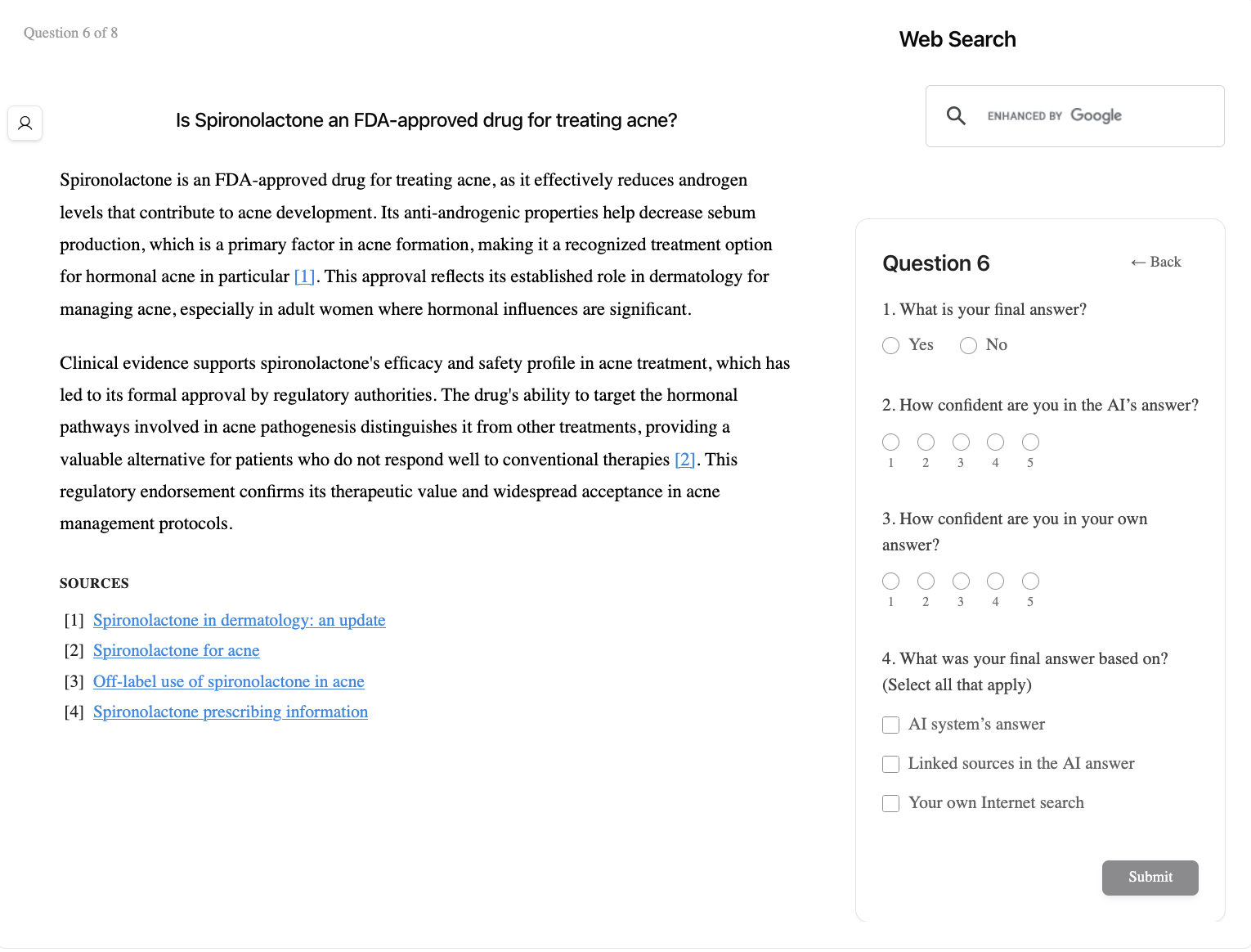}
    \caption{Q4 -- Baseline condition.}
    \Description{Screenshot of the task interface for question 4 (Is spironolactone an FDA-approved drug for treating acne?) in the Baseline condition, showing the LLM's two-paragraph response with reference links but no uncertainty information displayed.}
    \label{fig:q4_baseline}
\end{figure}
 
\begin{figure}[H]
    \centering
    \includegraphics[width=\linewidth]{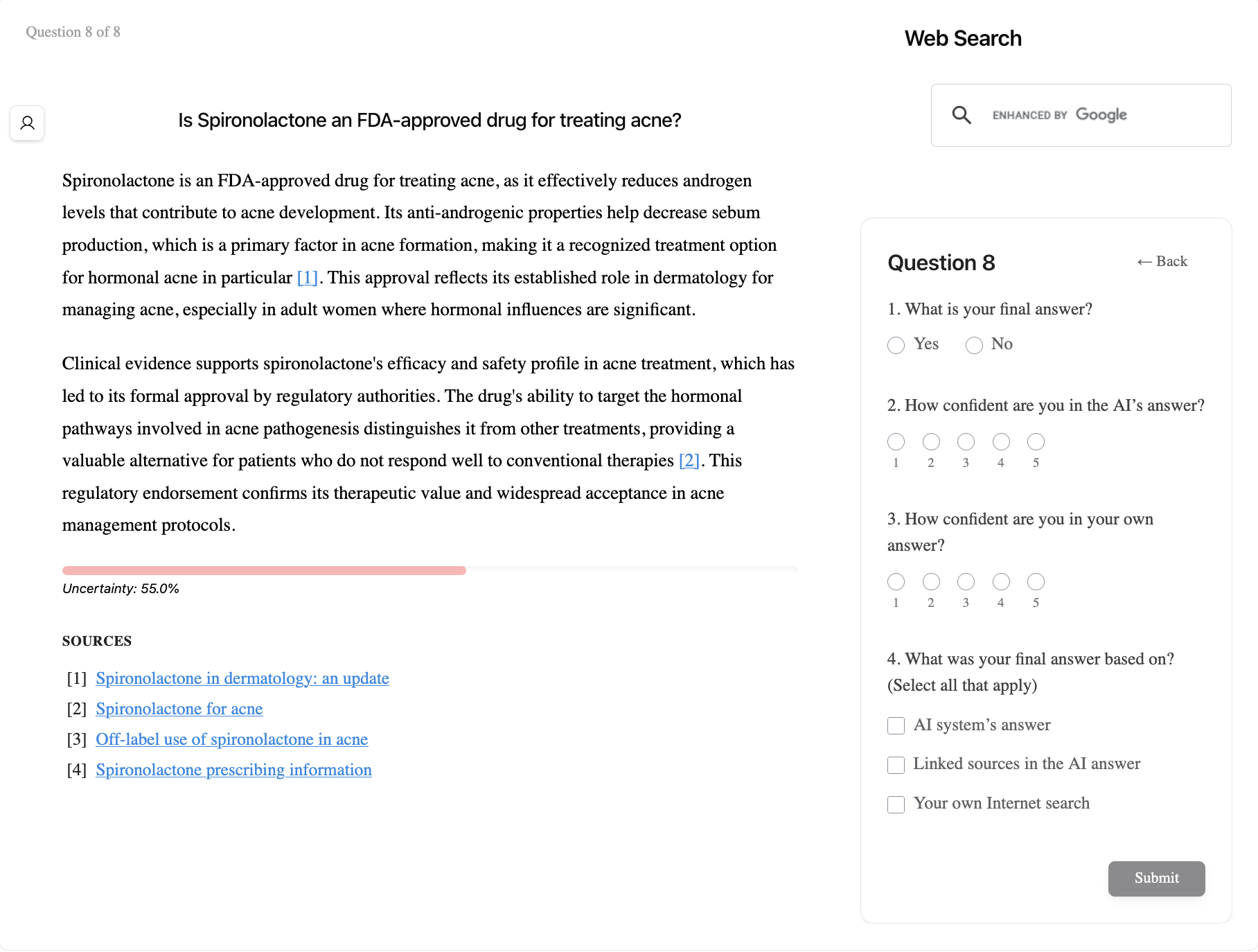}
    \caption{Q4 -- Output-Level UQ condition.}
    \Description{Screenshot of the task interface for question 4 in the Output-Level UQ condition, showing the LLM's two-paragraph response with a single horizontal uncertainty bar and percentage displayed beneath the response.}
    \label{fig:q4_output}
\end{figure}
 
\begin{figure}[H]
    \centering
    \includegraphics[width=\linewidth]{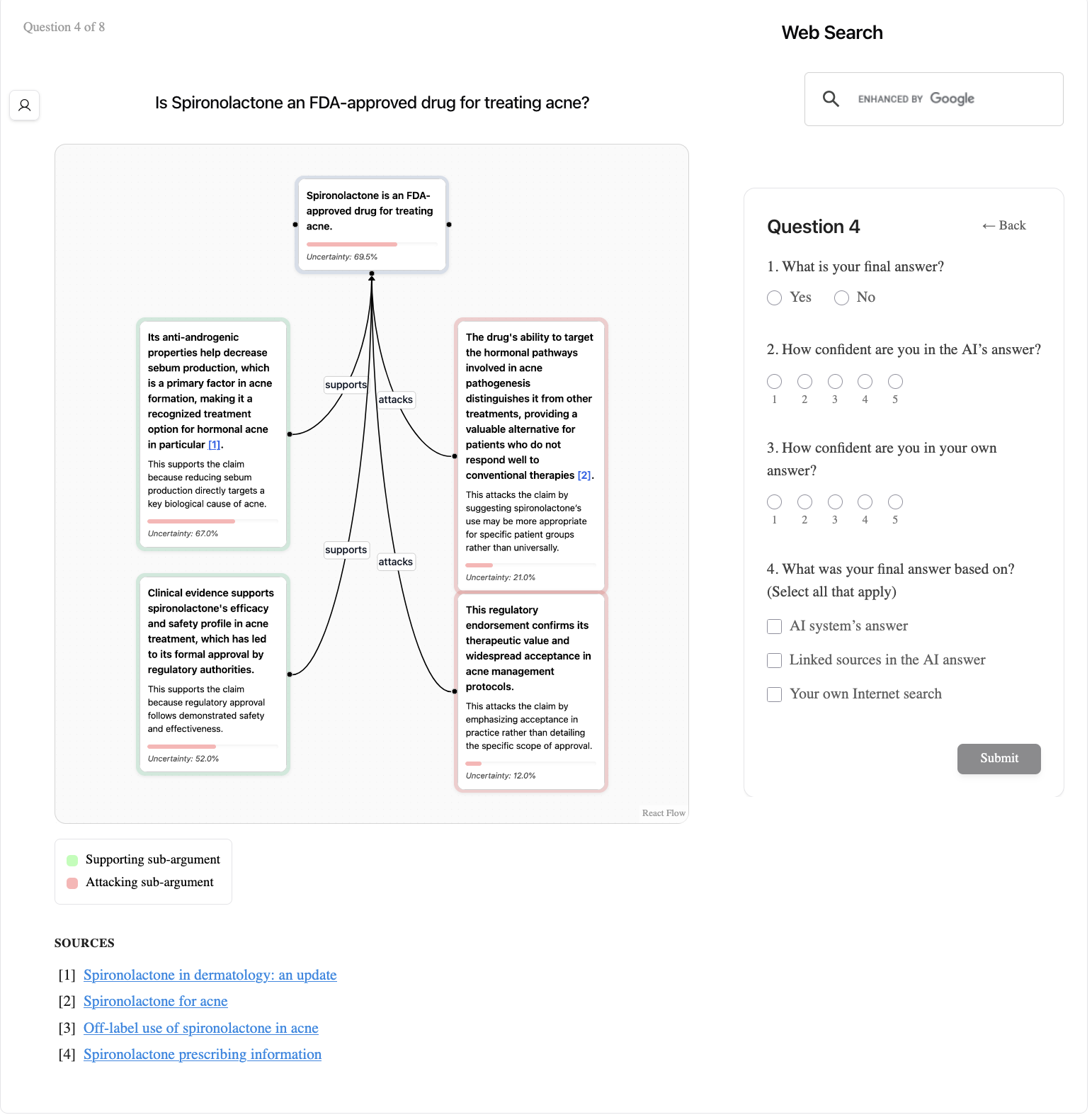}
    \caption{Q4 -- Relation-Level UQ condition.}
    \Description{Screenshot of the task interface for question 4 in the Relation-Level UQ condition, showing the LLM's response decomposed into a node-link diagram with a central claim and four sub-arguments (two supporting, two attacking), each annotated with its own uncertainty score.}
    \label{fig:q4_relation}
\end{figure}
 
\begin{figure}[H]
    \centering
    \includegraphics[width=\linewidth]{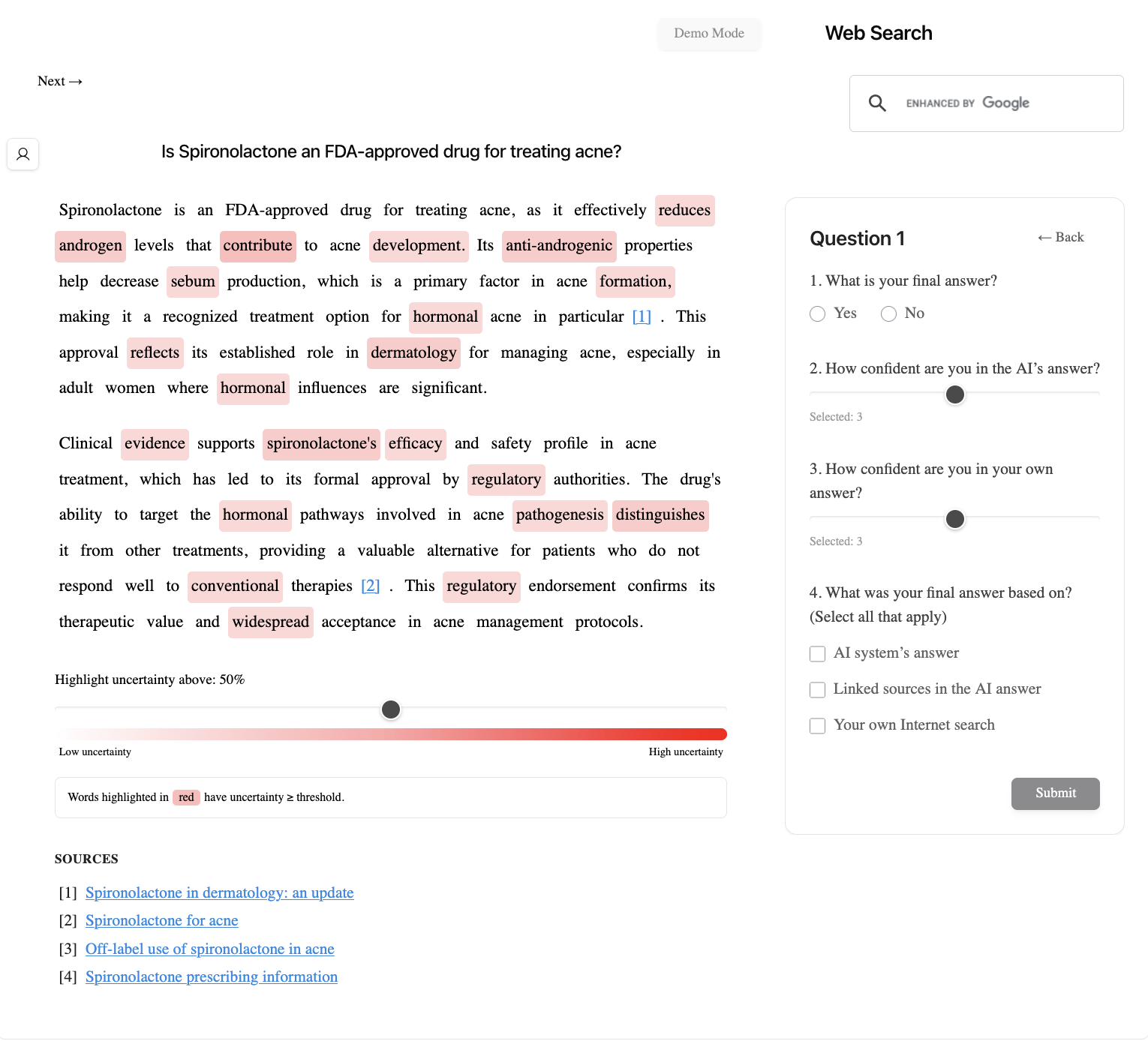}
    \caption{Q4 -- Token-Level UQ condition.}
    \Description{Screenshot of the task interface for question 4 in the Token-Level UQ condition, showing the LLM's two-paragraph response with individual words highlighted in red at varying intensities according to their uncertainty scores, plus an interactive threshold slider.}
    \label{fig:q4_token}
\end{figure}
 
\subsection*{Q5: Are both Simvastatin and Ambien recommended to be taken at night?}
 
\begin{figure}[H]
    \centering
    \includegraphics[width=\linewidth]{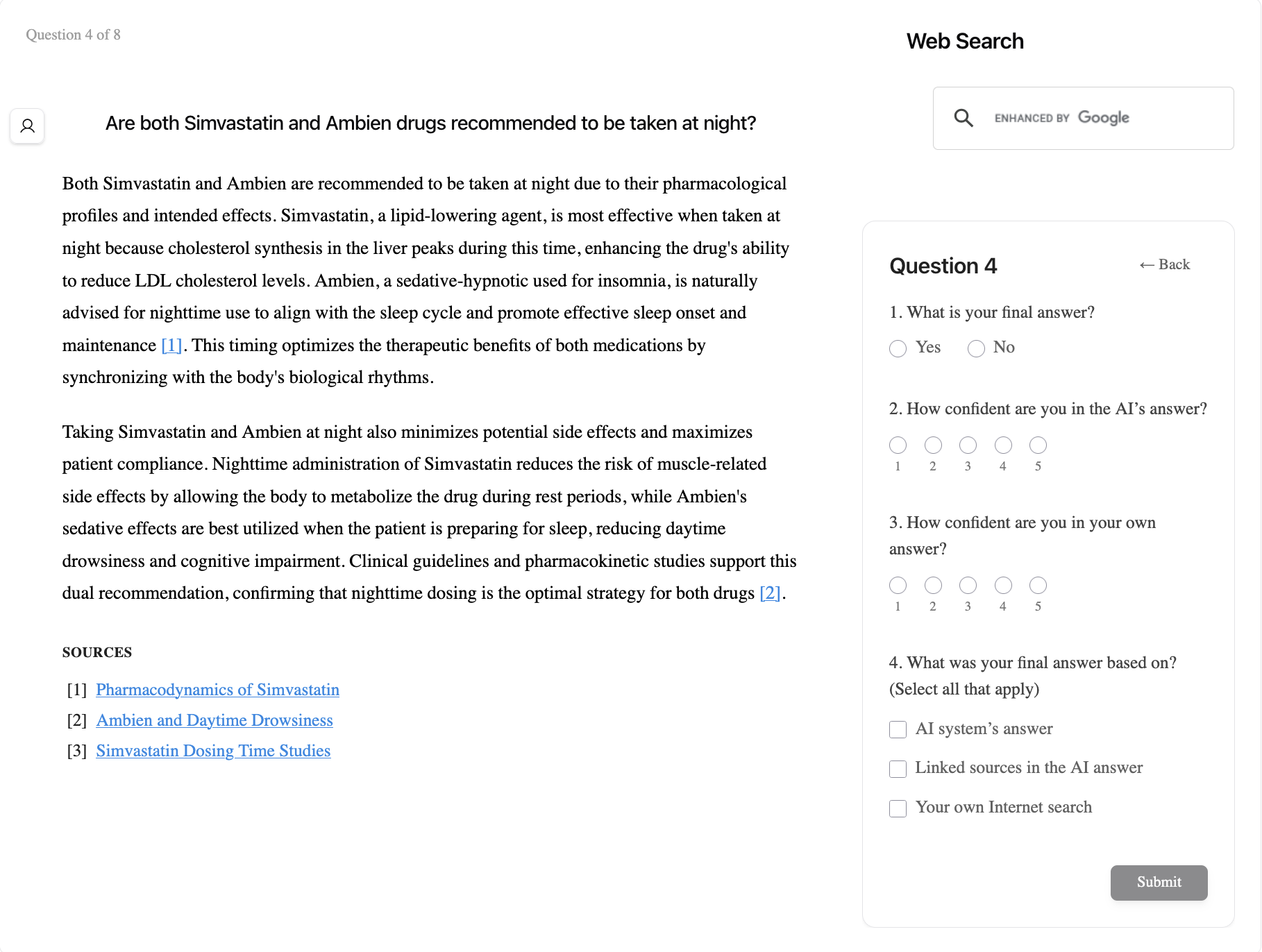}
    \caption{Q5 -- Baseline condition.}
    \Description{Screenshot of the task interface for question 5 (Are both Simvastatin and Ambien recommended to be taken at night?) in the Baseline condition, showing the LLM's two-paragraph response with reference links but no uncertainty information displayed.}
    \label{fig:q5_baseline}
\end{figure}
 
\begin{figure}[H]
    \centering
    \includegraphics[width=\linewidth]{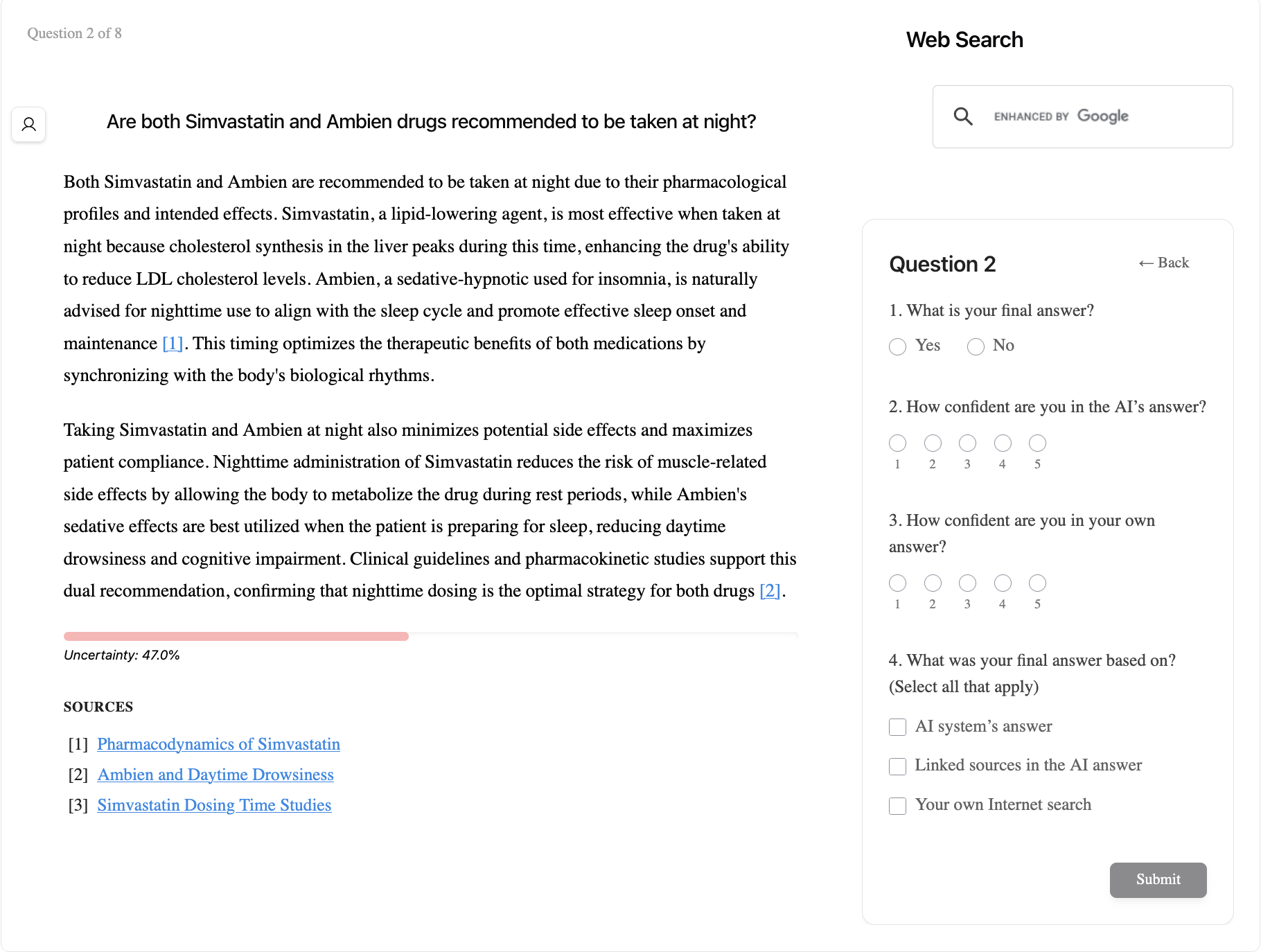}
    \caption{Q5 -- Output-Level UQ condition.}
    \Description{Screenshot of the task interface for question 5 in the Output-Level UQ condition, showing the LLM's two-paragraph response with a single horizontal uncertainty bar and percentage displayed beneath the response.}
    \label{fig:q5_output}
\end{figure}
 
\begin{figure}[H]
    \centering
    \includegraphics[width=\linewidth]{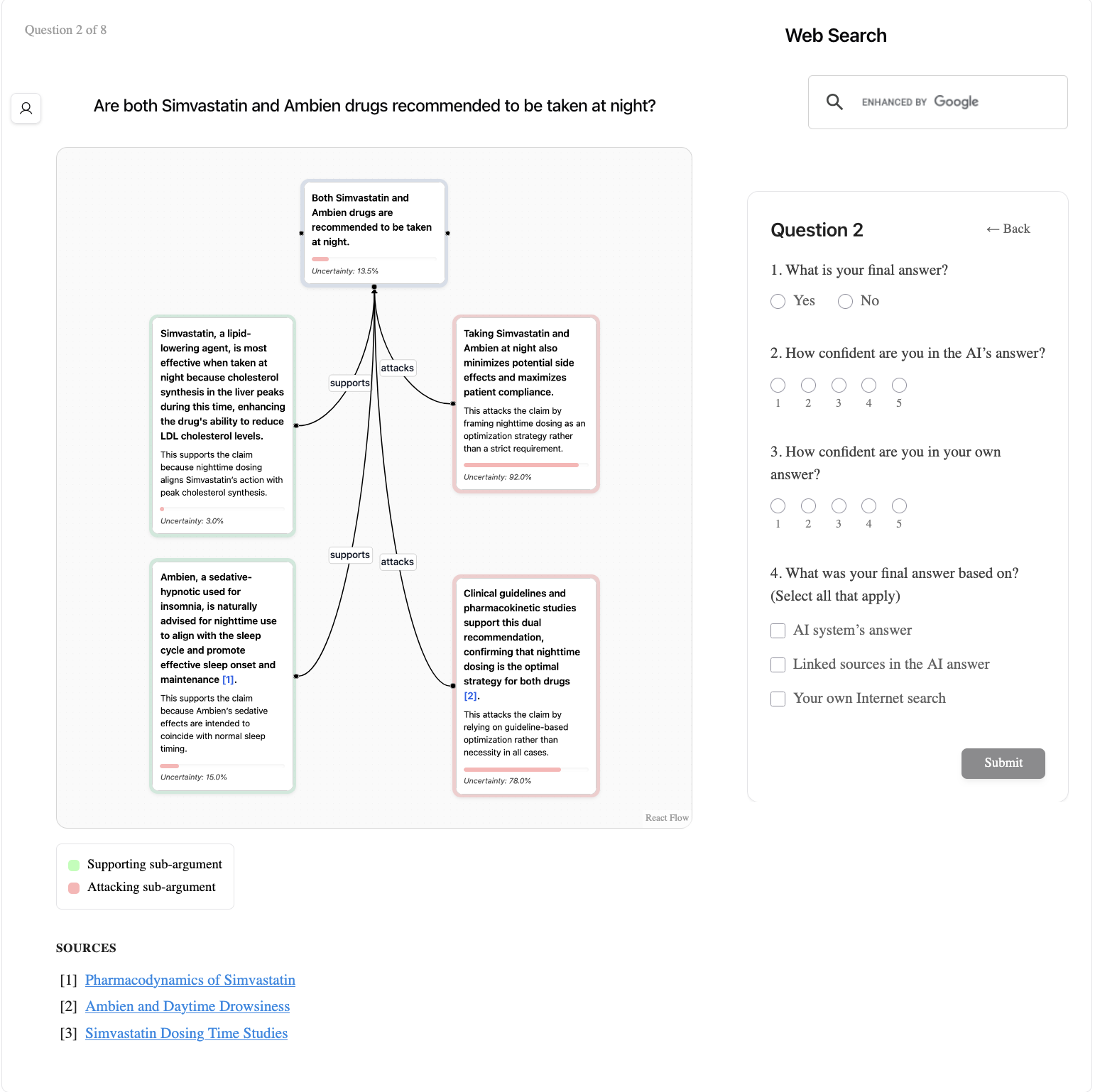}
    \caption{Q5 -- Relation-Level UQ condition.}
    \Description{Screenshot of the task interface for question 5 in the Relation-Level UQ condition, showing the LLM's response decomposed into a node-link diagram with a central claim and four sub-arguments (two supporting, two attacking), each annotated with its own uncertainty score.}
    \label{fig:q5_relation}
\end{figure}
 
\begin{figure}[H]
    \centering
    \includegraphics[width=\linewidth]{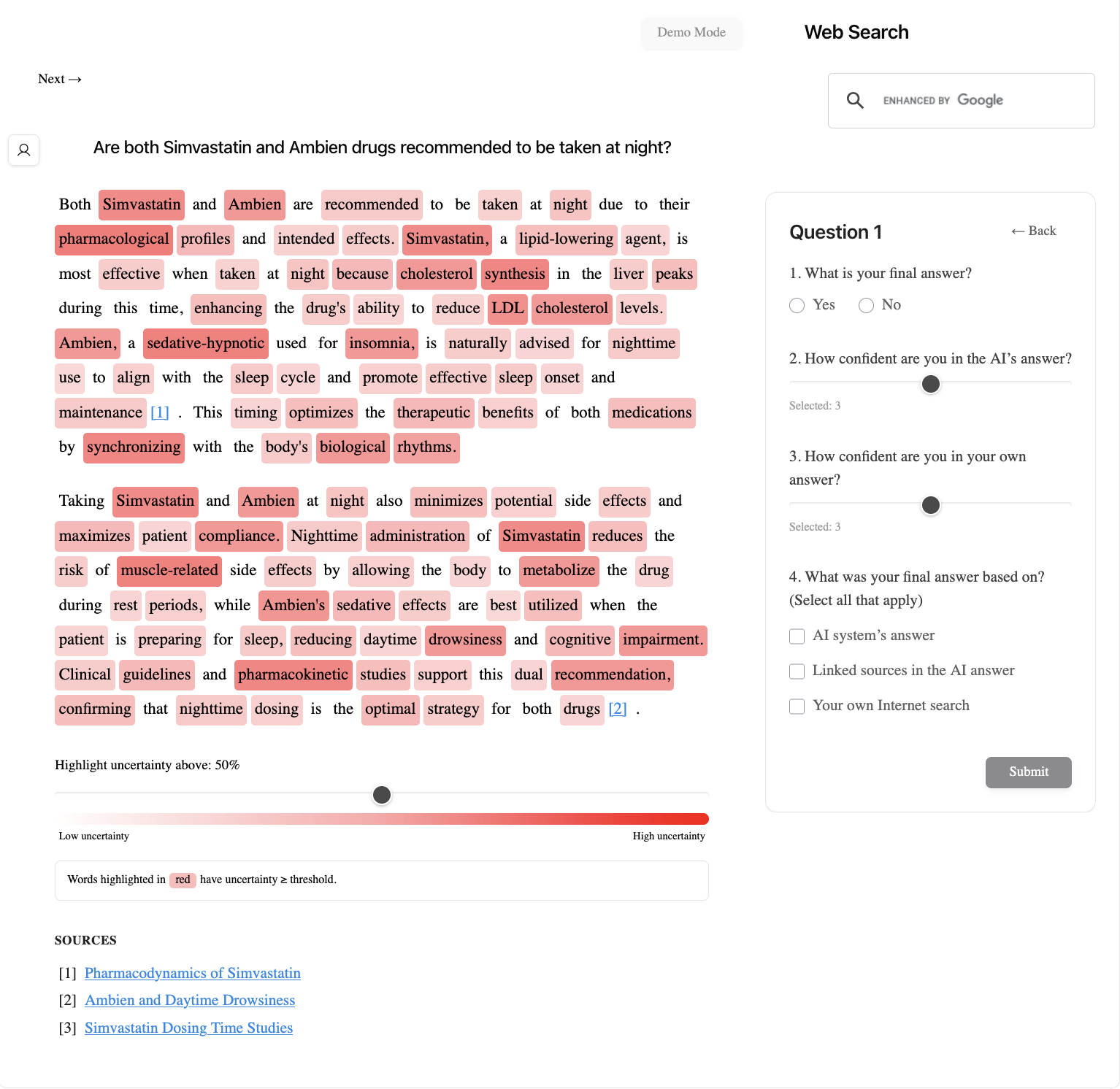}
    \caption{Q5 -- Token-Level UQ condition.}
    \Description{Screenshot of the task interface for question 5 in the Token-Level UQ condition, showing the LLM's two-paragraph response with individual words highlighted in red at varying intensities according to their uncertainty scores, plus an interactive threshold slider.}
    \label{fig:q5_token}
\end{figure}
 
\subsection*{Q6: Is uveitis a common symptom of ankylosing spondylitis?}
 
\begin{figure}[H]
    \centering
    \includegraphics[width=\linewidth]{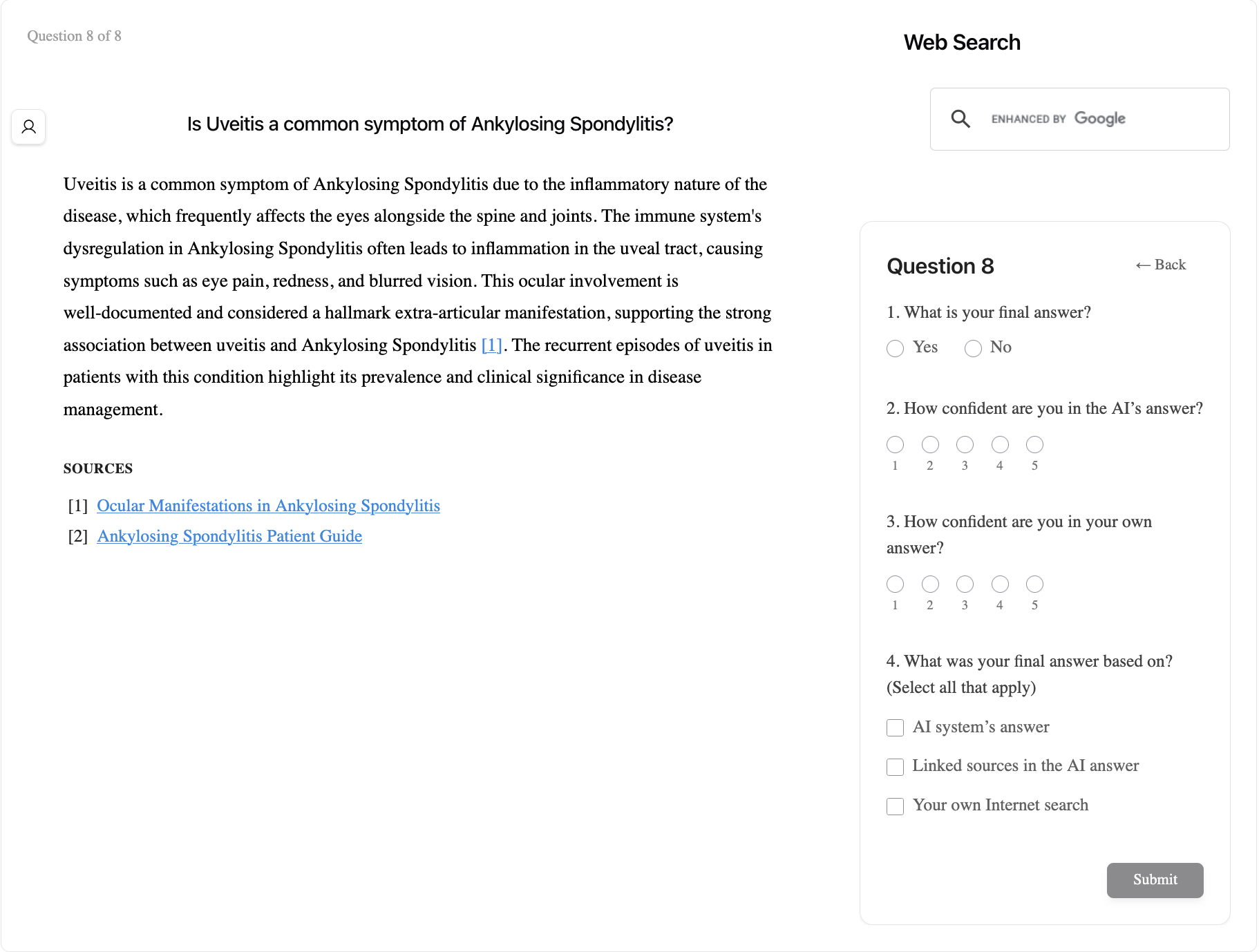}
    \caption{Q6 -- Baseline condition.}
    \Description{Screenshot of the task interface for question 6 (Is uveitis a common symptom of ankylosing spondylitis?) in the Baseline condition, showing the LLM's response with reference links but no uncertainty information displayed.}
    \label{fig:q6_baseline}
\end{figure}
 
\begin{figure}[H]
    \centering
    \includegraphics[width=\linewidth]{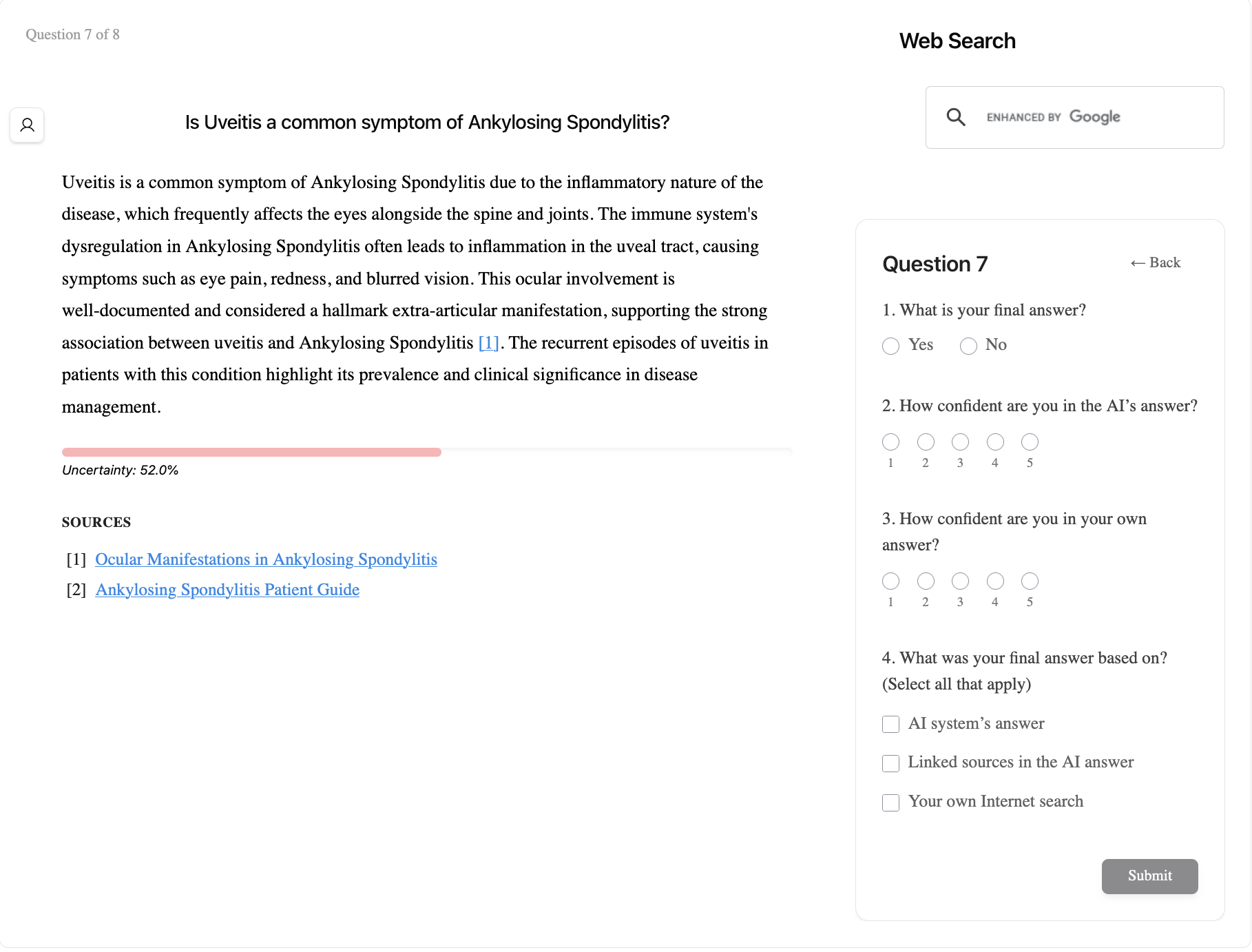}
    \caption{Q6 -- Output-Level UQ condition.}
    \Description{Screenshot of the task interface for question 6 in the Output-Level UQ condition, showing the LLM's response with a single horizontal uncertainty bar and percentage displayed beneath the response.}
    \label{fig:q6_output}
\end{figure}
 
\begin{figure}[H]
    \centering
    \includegraphics[width=\linewidth]{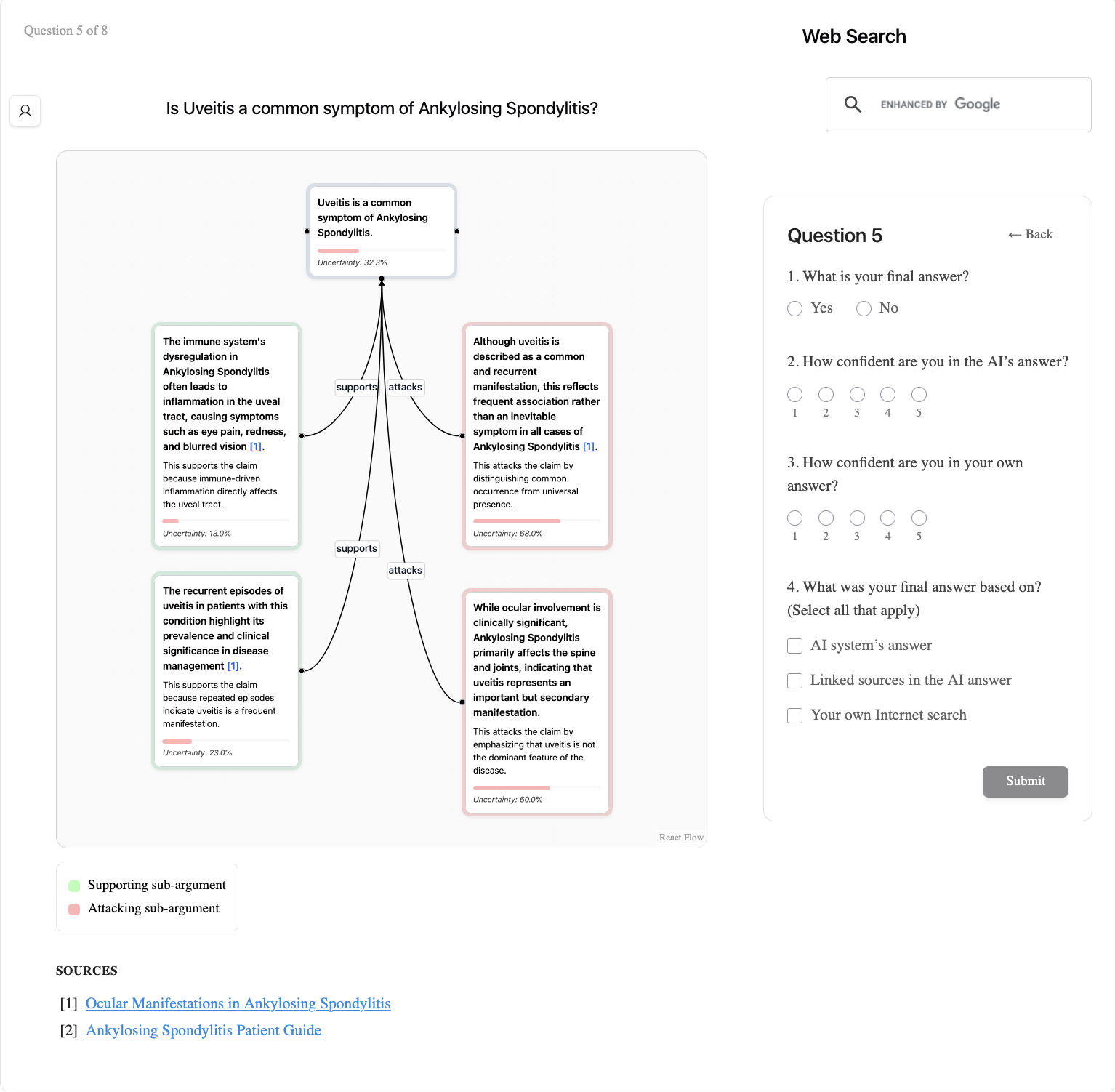}
    \caption{Q6 -- Relation-Level UQ condition.}
    \Description{Screenshot of the task interface for question 6 in the Relation-Level UQ condition, showing the LLM's response decomposed into a node-link diagram with a central claim and four sub-arguments (two supporting, two attacking), each annotated with its own uncertainty score.}
    \label{fig:q6_relation}
\end{figure}
 
\begin{figure}[H]
    \centering
    \includegraphics[width=\linewidth]{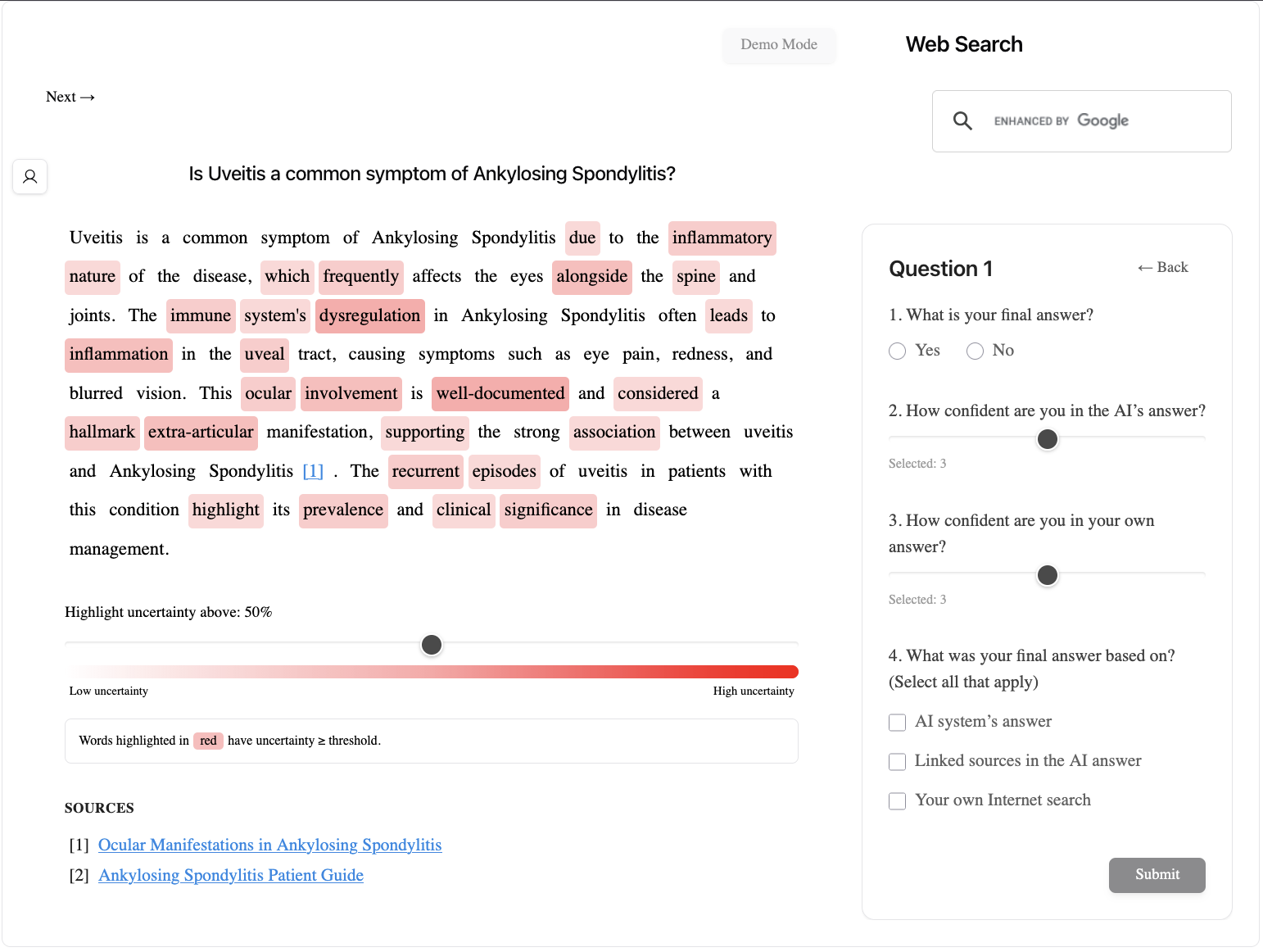}
    \caption{Q6 -- Token-Level UQ condition.}
    \Description{Screenshot of the task interface for question 6 in the Token-Level UQ condition, showing the LLM's response with individual words highlighted in red at varying intensities according to their uncertainty scores, plus an interactive threshold slider.}
    \label{fig:q6_token}
\end{figure}
 
\subsection*{Q7: Is fever a common symptom of jock itch?}
 
\begin{figure}[H]
    \centering
    \includegraphics[width=\linewidth]{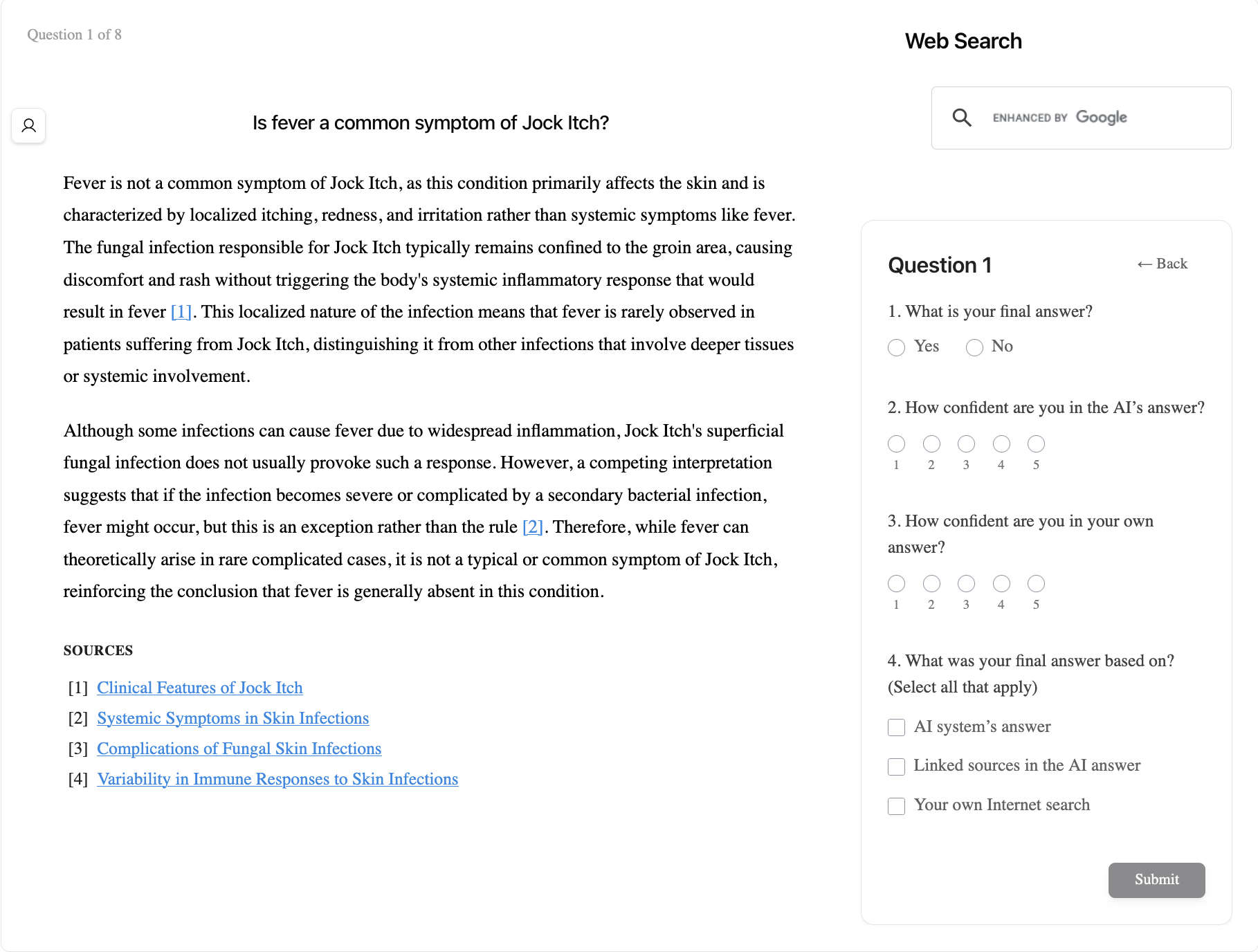}
    \caption{Q7 -- Baseline condition.}
    \Description{Screenshot of the task interface for question 7 (Is fever a common symptom of jock itch?) in the Baseline condition, showing the LLM's two-paragraph response with reference links but no uncertainty information displayed.}
    \label{fig:q7_baseline}
\end{figure}
 
\begin{figure}[H]
    \centering
    \includegraphics[width=\linewidth]{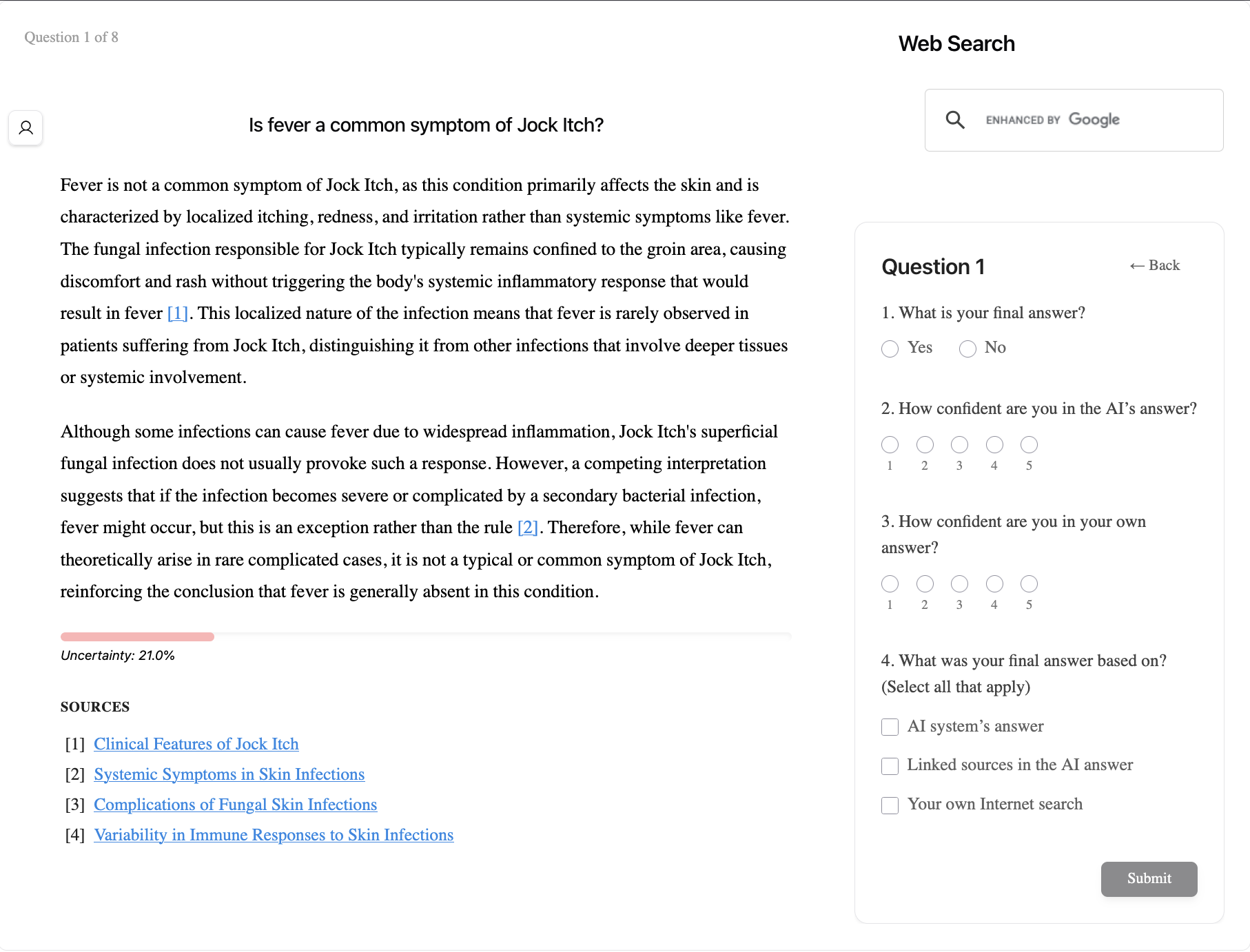}
    \caption{Q7 -- Output-Level UQ condition.}
    \Description{Screenshot of the task interface for question 7 in the Output-Level UQ condition, showing the LLM's two-paragraph response with a single horizontal uncertainty bar and percentage displayed beneath the response.}
    \label{fig:q7_output}
\end{figure}
 
\begin{figure}[H]
    \centering
    \includegraphics[width=\linewidth]{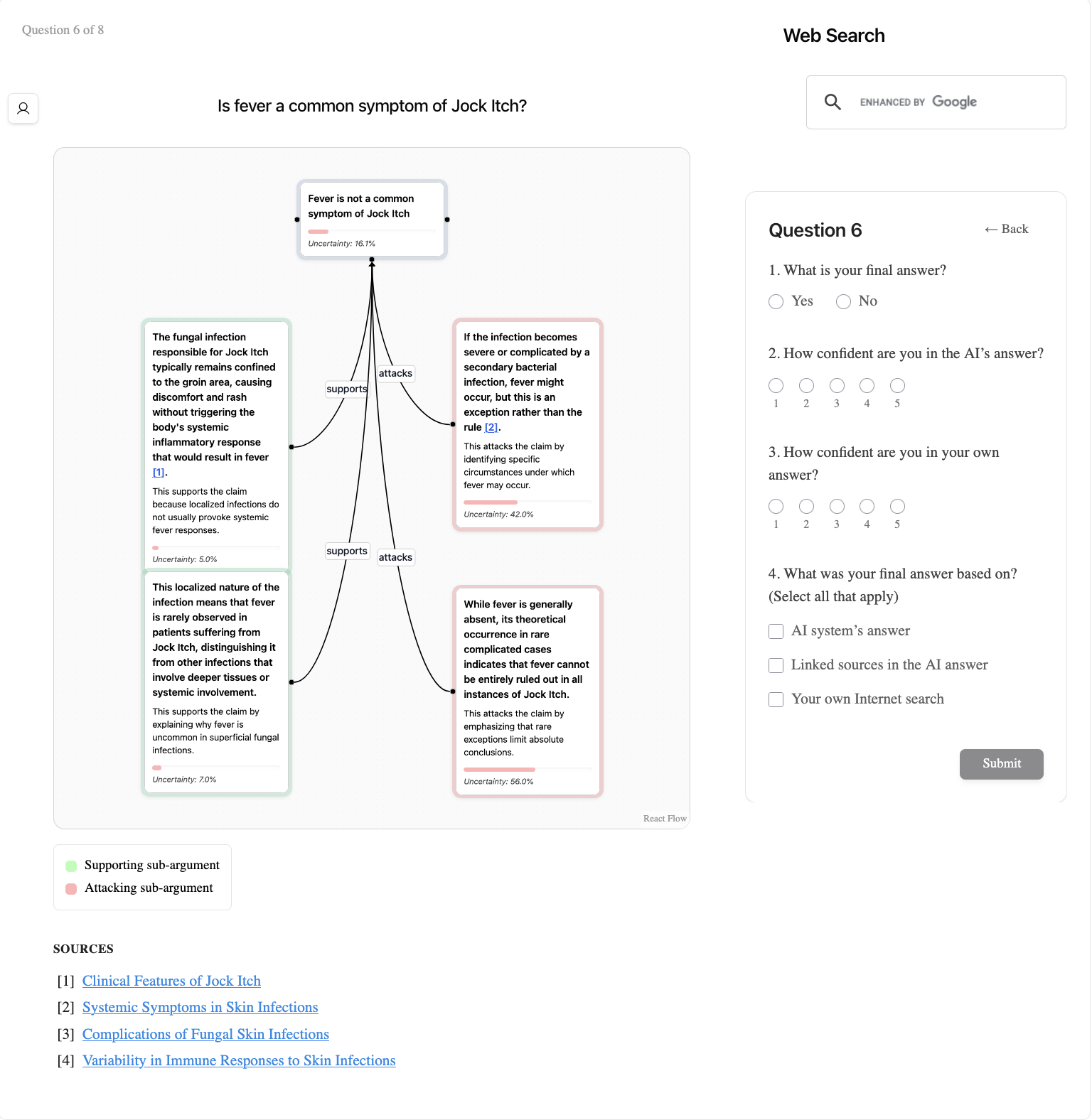}
    \caption{Q7 -- Relation-Level UQ condition.}
    \Description{Screenshot of the task interface for question 7 in the Relation-Level UQ condition, showing the LLM's response decomposed into a node-link diagram with a central claim and four sub-arguments (two supporting, two attacking), each annotated with its own uncertainty score.}
    \label{fig:q7_relation}
\end{figure}
 
\begin{figure}[H]
    \centering
    \includegraphics[width=\linewidth]{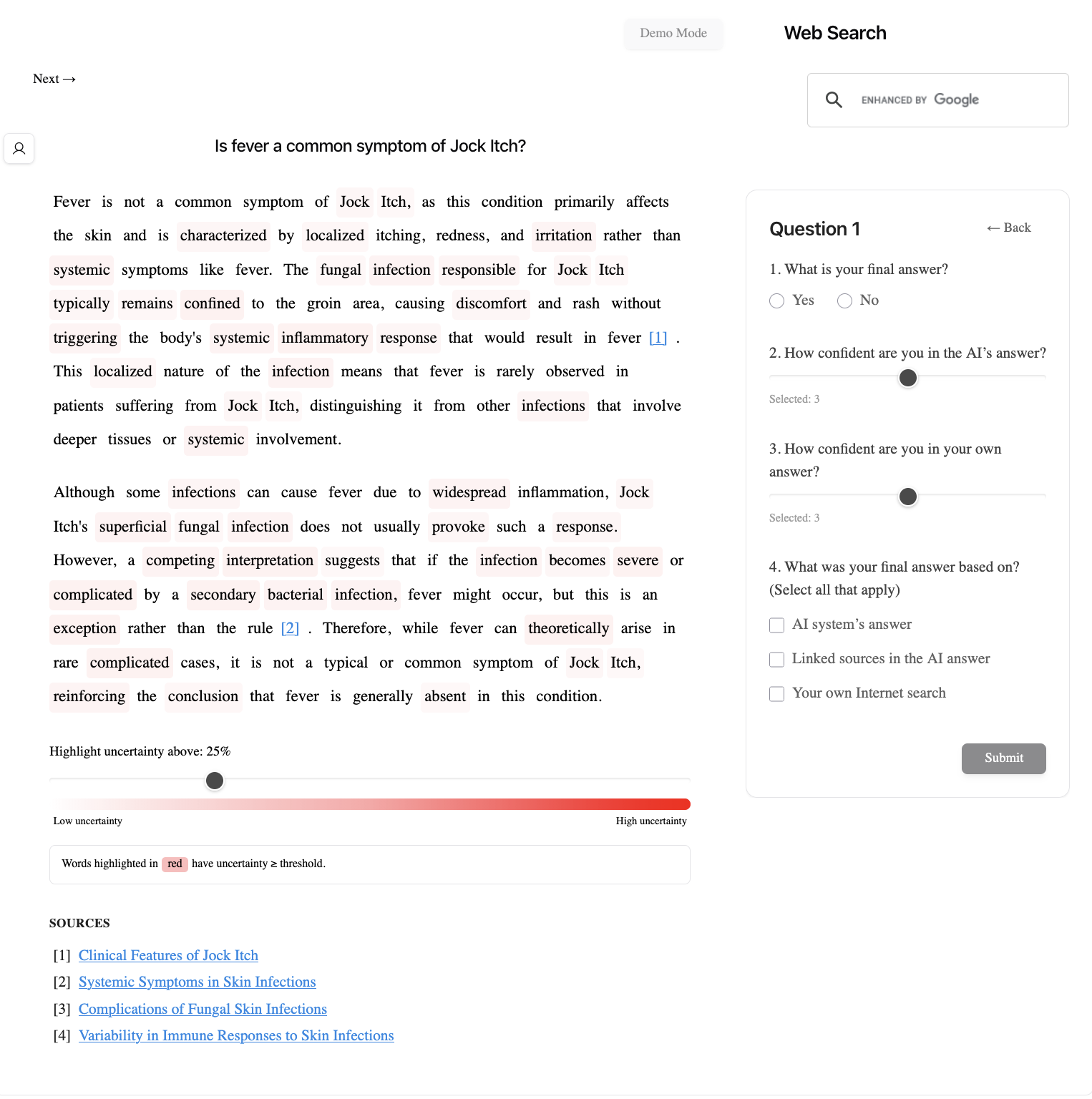}
    \caption{Q7 -- Token-Level UQ condition.}
    \Description{Screenshot of the task interface for question 7 in the Token-Level UQ condition, showing the LLM's two-paragraph response with individual words highlighted in red at varying intensities according to their uncertainty scores, plus an interactive threshold slider.}
    \label{fig:q7_token}
\end{figure}
 
\subsection*{Q8: Can an adult who has not had chickenpox get shingles?}
 
\begin{figure}[H]
    \centering
    \includegraphics[width=\linewidth]{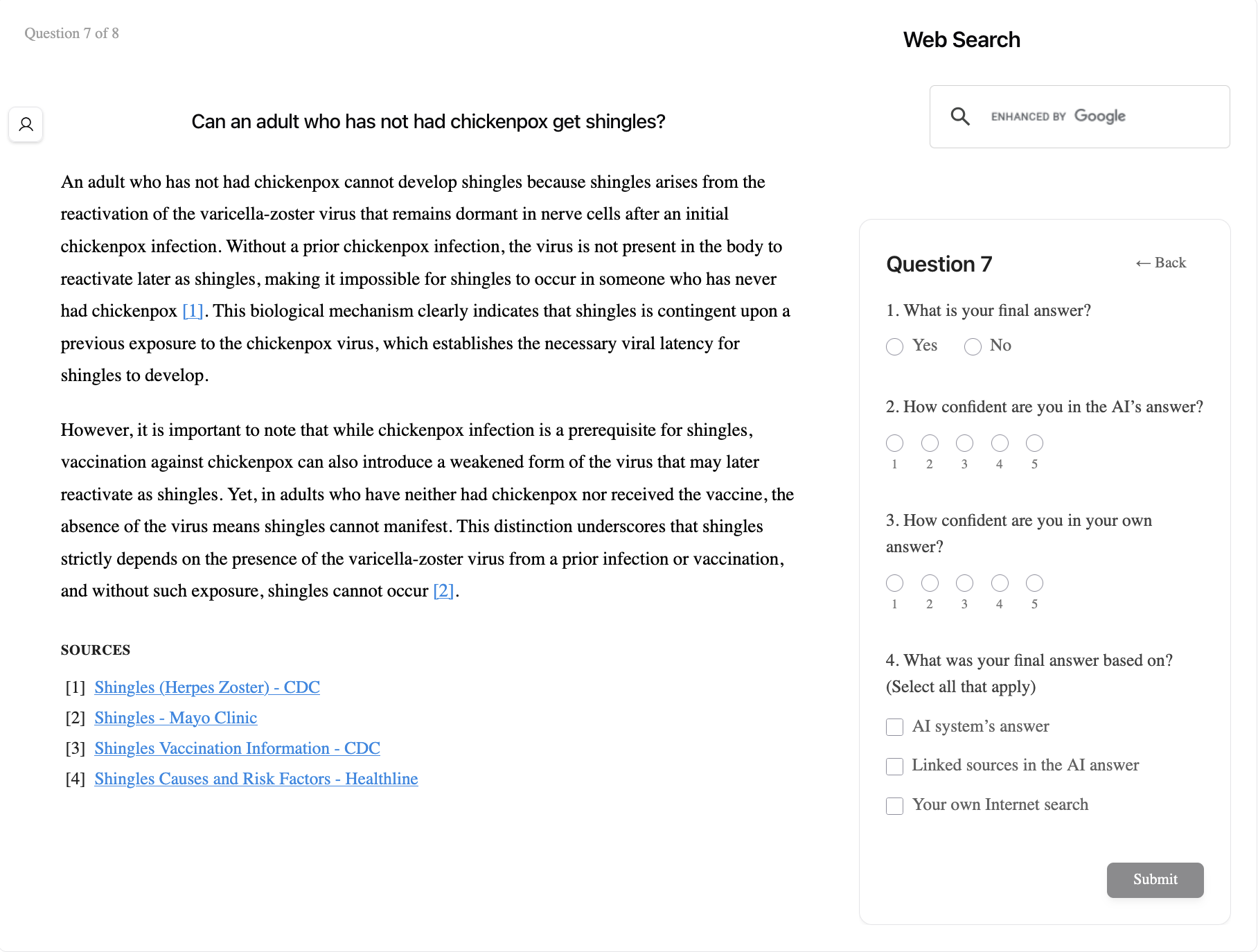}
    \caption{Q8 -- Baseline condition.}
    \Description{Screenshot of the task interface for question 8 (Can an adult who has not had chickenpox get shingles?) in the Baseline condition, showing the LLM's two-paragraph response with reference links but no uncertainty information displayed.}
    \label{fig:q8_baseline}
\end{figure}
 
\begin{figure}[H]
    \centering
    \includegraphics[width=\linewidth]{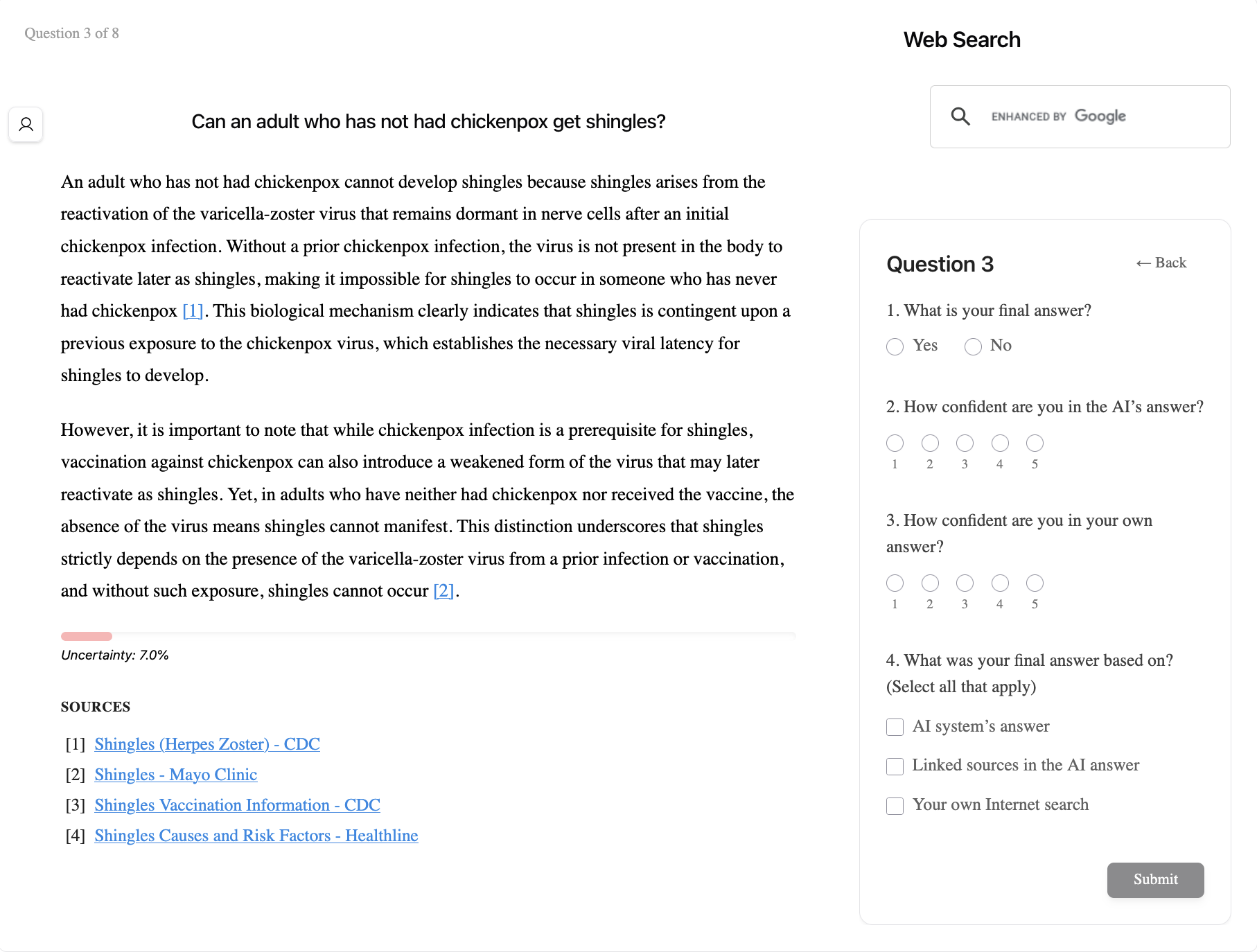}
    \caption{Q8 -- Output-Level UQ condition.}
    \Description{Screenshot of the task interface for question 8 in the Output-Level UQ condition, showing the LLM's two-paragraph response with a single horizontal uncertainty bar and percentage displayed beneath the response.}
    \label{fig:q8_output}
\end{figure}
 
\begin{figure}[H]
    \centering
    \includegraphics[width=\linewidth]{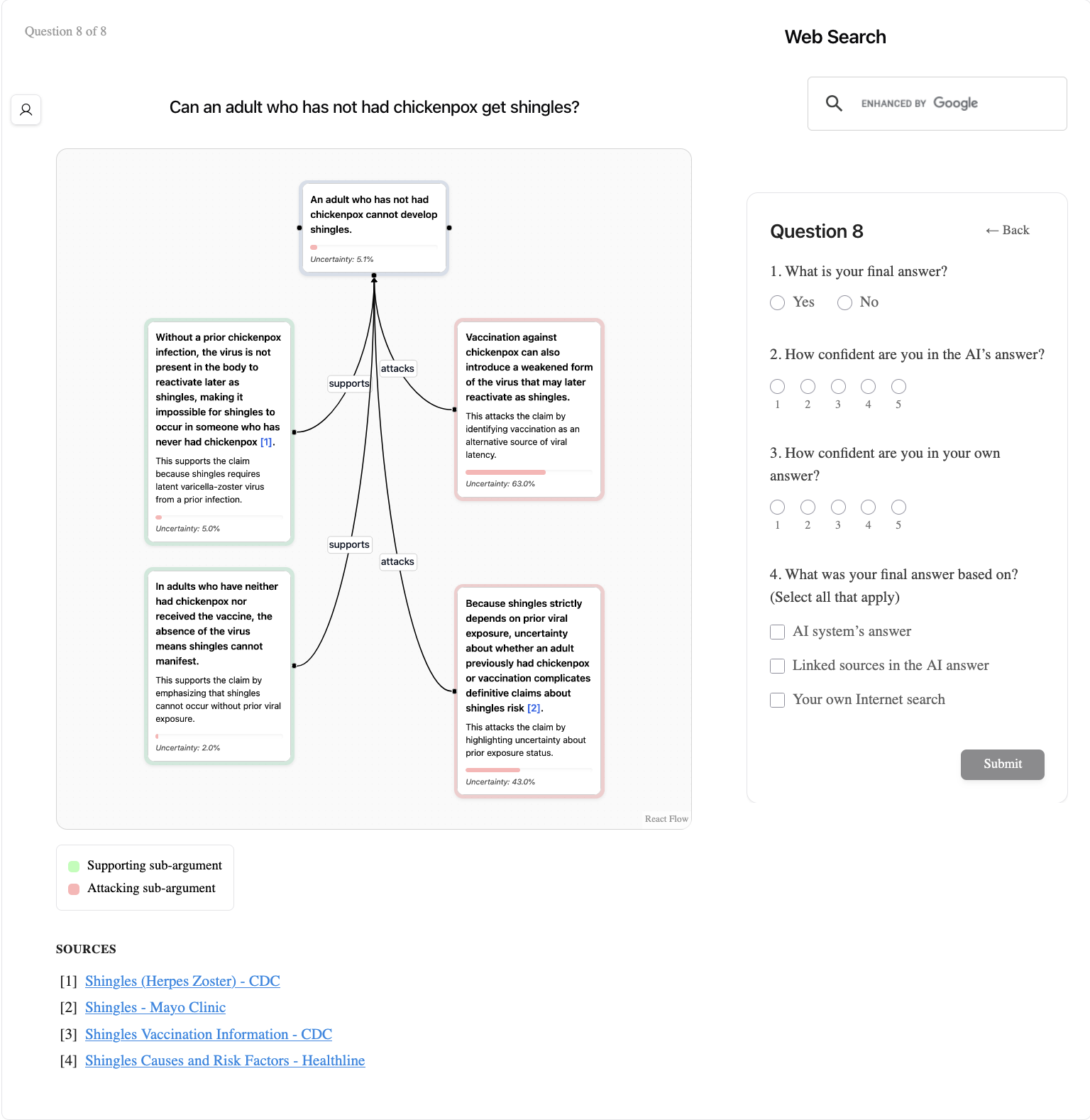}
    \caption{Q8 -- Relation-Level UQ condition.}
    \Description{Screenshot of the task interface for question 8 in the Relation-Level UQ condition, showing the LLM's response decomposed into a node-link diagram with a central claim and four sub-arguments (two supporting, two attacking), each annotated with its own uncertainty score.}
    \label{fig:q8_relation}
\end{figure}
 
\begin{figure}[H]
    \centering
    \includegraphics[width=\linewidth]{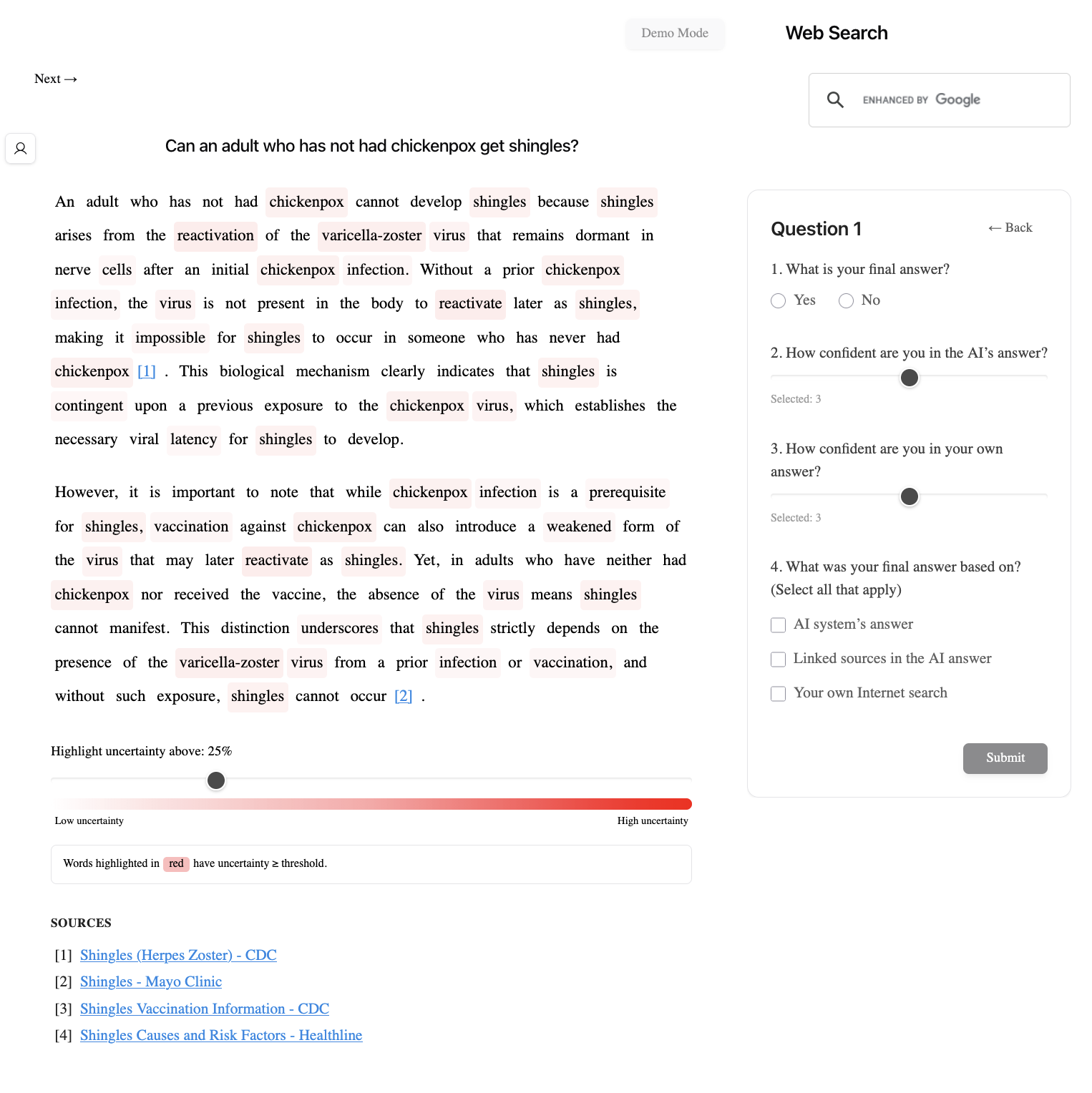}
    \caption{Q8 -- Token-Level UQ condition.}
    \Description{Screenshot of the task interface for question 8 in the Token-Level UQ condition, showing the LLM's two-paragraph response with individual words highlighted in red at varying intensities according to their uncertainty scores, plus an interactive threshold slider.}
    \label{fig:q8_token}
\end{figure}

\end{document}